\documentclass[pdflatex,sn-mathphys,Numbered]{sn-jnl}

\usepackage{graphicx}%
\usepackage{amsmath,amssymb,amsfonts}%
\usepackage{booktabs}%
\usepackage{xcolor}%
\usepackage{textcomp}%
\usepackage{manyfoot}%
\usepackage{placeins}%
\usepackage{float}%
\usepackage{longtable}%

\newcommand{\tablefontsize}{\small}

\begin{document}

\title[TiN coating grain boundary degradation]{Grain boundary strength and corrosion cracking susceptibility in TiN coatings under fission product segregation}

\author*[1]{\fnm{Jia-Hong} \sur{Ke}}\email{jiahong.ke@inl.gov}
\author*[2]{\fnm{Rongjie} \sur{Song}}\email{rongjie.song@inl.gov}

\affil*[1]{\orgdiv{Computational Mechanics and Materials Department}, \orgname{Idaho National Laboratory}, \orgaddress{\city{Idaho Falls}, \postcode{83415}, \state{ID}, \country{USA}}}
\affil*[2]{\orgdiv{Irradiated Fuels and Materials Department}, \orgname{Idaho National Laboratory}, \orgaddress{\city{Idaho Falls}, \postcode{83415}, \state{ID}, \country{USA}}}

\abstract{During fuel--cladding chemical interaction (FCCI), lanthanide fission products generated in the fuel can migrate into cladding grain boundaries (GBs), where segregation may reduce GB strength and promote corrosion cracking. We use first-principles segregation, decohesion, and bonding calculations to quantify Nd and Ce effects on six coincident-site lattice GBs in bcc Fe and TiN. For each material, segregation energies, work-of-separation values for complete interfacial cleavage, and traction--separation curves measuring ideal resistance to crack opening were evaluated within a consistent computational framework. Nd and Ce segregate favorably to both materials, and the median segregation energies are more negative in TiN, showing that TiN does not simply exclude these species from internal interfaces. The mechanical responses of Fe and TiN differ sharply. In Fe, Nd and Ce consistently reduce work of separation and peak traction, with median work-of-separation retentions of 0.72 and 0.65 and median peak-traction retentions of 0.47 and 0.43. In TiN, most decorated GBs preserve much of the clean response, with median work-of-separation retentions of 1.00 and 0.97 and median peak-traction retention of 0.85 for both fission products. Bonding analysis shows weak, occupied antibonding X--Fe replacement bonds in Fe, whereas the surrounding Ti--N network remains largely preserved after Nd/Ce segregation in TiN. These results provide atomistic support for the use of TiN and related nitride coatings to reduce FCCI damage by retaining GB strength if fission products segregate to coating internal interfaces. This retention of GB strength also provides a mechanistic descriptor for assessing coating susceptibility to FCCI-assisted corrosion cracking and environmentally assisted intergranular fracture.}


\maketitle

\section{Introduction}

The long-term reliability of fuel--cladding systems in nuclear environments can be limited by chemical attack and mechanical weakening at internal interfaces. During fuel--cladding chemical interaction (FCCI), fission products generated in the fuel can migrate into adjacent cladding materials and accumulate at grain boundaries (GBs). Because these species are intrinsic products of the fission process, mitigation must limit their migration and reduce the damage they cause after reaching internal interfaces. GBs are efficient sinks for irradiation defects and mobile chemical species because they contain excess free volume, geometrical disorder, and disrupted bonding relative to the bulk lattice \cite{lejcek1995thermodynamics,cantwell2014grain,hu2024computational}. Lanthanide infiltration is particularly concerning because it can promote intergranular embrittlement, stress corrosion cracking, and premature failure in nuclear environments \cite{keiser2019fuel,mariani2011lanthanides,keiser20123,matthews2017fuel}. FCCI-driven GB weakening can therefore be viewed as environmentally assisted fracture, where radiation-generated chemical species from the service environment change the mechanical resistance of internal interfaces. Thus, understanding how fission products change GB strength is needed to connect fission-product migration into cladding with the formation of mechanically vulnerable paths for environmentally assisted intergranular fracture and corrosion cracking.

Nd and Ce are representative lanthanide fission products because their atomic sizes and electronic structures differ strongly from those of the transition-metal and ceramic materials considered here. These elements have been reported to migrate toward internal interfaces in nuclear fuel and cladding materials \cite{matthews2017fuel}. Once incorporated at GB sites, they can modify local bonding and reduce interfacial strength \cite{seah1975interface,lejcek2017interfacial,gibson2015segregation}. Recent work on HT9 ferritic--martensitic cladding showed that fission-product infiltration can drive GB corrosion cracking, linking fission-product chemistry with environmentally assisted intergranular fracture \cite{wang2022transmission,salvato2024effect,harp2017scanning}. Therefore, the effects of Nd and Ce on GB strength need to be evaluated directly.

The effect of segregated fission products on GB cohesion is usually described by the thermodynamic model developed by Rice and Wang \cite{rice1989embrittlement,gibson2015segregation}. In this model, embrittlement occurs when a segregated solute stabilizes the separated fracture surfaces more than the intact GB, thereby reducing the work required for interfacial separation. First-principles studies have connected this thermodynamic effect to charge redistribution, bond weakening, and modified hybridization across the interface \cite{messmer1982role,wu1994first,wu1994nature,geng2001influence,yamaguchi2005grain}. Large-scale surveys of \textit{ab initio} segregation data show that segregation-induced embrittlement is often captured by bond-breaking arguments \cite{gibson2016survey}. At the same time, dopant studies show that chemistry and GB site preference can lead to deviations from simple bond-breaking trends \cite{tran2016computational}. Thus, favorable segregation alone does not determine whether a GB will become a mechanically weak interface.

Density functional theory (DFT) combined with first-principles cohesive-zone analysis provides a direct way to quantify this degradation step. Segregation energies \(E_{\mathrm{seg}}\), works of separation \(W_{\mathrm{sep}}\), and traction--separation relationships can be obtained from the same atomistic description \cite{van2004thermodynamics,enrique2017decohesion}. The work of separation gives the energy per area required for complete interfacial separation into fracture surfaces, whereas the peak traction \(\sigma_{\max}\) is the maximum cohesive stress along the ideal opening path and provides an atomistic measure of resistance to the onset of local separation. These metrics distinguish fission products that segregate to GBs from those that also cause severe decohesion, which is the atomistic precursor to fission-product-assisted intergranular damage. Such calculations do not replace transport, corrosion-kinetics, or crack-growth models. They provide the atomistic input needed to determine whether a segregated interface is likely to become a mechanically weak path.

Body-centered cubic (bcc) Fe is an appropriate baseline for this problem because ferritic and ferritic--martensitic Fe-based alloys are widely used in cladding applications and are sensitive to radiation-induced chemical redistribution \cite{ke2023microstructure,ke2022cluster,zheng2019correlation,ke2018cluster}. In the present comparison, bcc Fe provides a tractable metallic reference for evaluating how fission products modify GB strength in Fe-based cladding. Recent high-throughput first-principles work by Mai et al. established reference data for clean and solute-decorated bcc Fe GB strength across many boundary characters \cite{mai2025high}. However, the specific effects of Nd and Ce, and the comparison with candidate coating materials under identical segregation and decohesion metrics, remain less developed.

Transition-metal nitride coatings offer a possible route to reducing fission-product damage in Fe-based cladding. TiN is attractive because of its thermal stability, corrosion resistance, and mixed metallic--ionic--covalent bonding, which combine ceramic-like hardness with metallic conductivity \cite{toth2014transition,yu2015phase,blaha1985bonding,sundgren1985structure}. Atomistic studies of TiN GBs have shown that boundary character strongly affects GB energy, point-defect behavior, oxidation, and impurity transport \cite{mckenna2018structure,triestram2024hydride}. In general, solute decoration can change GB energy, mobility, strength, and structure \cite{raabe2014segregation}. TiN GBs may therefore be chemically active interfaces rather than inert barriers. Recent ion-irradiation and diffusion-couple studies showed that TiN coatings suppress Ce--Fe interaction under conditions where uncoated Fe-based systems react strongly \cite{khatkhatay2013diffusion,gan2023status,gan2024effect,gan2024summary}. These studies attributed the protective effect mainly to TiN acting as a diffusion barrier and limiting the transport of Ce and Fe through the coating.

These studies clarify fission-product transport through TiN, but they do not answer what happens if fission products reach TiN internal interfaces, where GB diffusion and segregation can occur. In that case, the degradation response depends on whether segregation also reduces the work required for interfacial separation and the peak traction sustained by the GB. Compared with metallic Fe, the more directional bonding in TiN may make GB strength less sensitive to fission-product-induced changes in local bonding. Given this uncertainty, a controlled comparison between bcc Fe and TiN can clarify whether TiN changes only fission-product access to interfaces, or also changes the strength penalty after segregation.

In the present study, we quantify the effect of Nd and Ce segregation on GB strength in bcc Fe and TiN using first-principles calculations. Segregation energies, GB excess energies, works of separation, and traction--separation relationships are evaluated for six coincident-site lattice (CSL) GBs per material. The calculation results are used to separate fission-product segregation from the strength penalty after segregation, and both quantities are related to local bonding changes. This comparison isolates one mechanistic step in fission-product-assisted intergranular degradation and provides atomistic metrics for evaluating coating GB degradation.

\section{Results}
The calculation results are presented in the following order. We first establish clean-interface baselines for Fe and TiN, then compare Nd/Ce segregation and the strength response after segregation. Clean GBs define reference values of the GB excess energy \(\gamma_{\mathrm{GB}}\), work of separation \(W_{\mathrm{sep}}\), and peak traction \(\sigma_{\max}\). Here \(W_{\mathrm{sep}}\) measures the energy scale for full GB cleavage, while \(\sigma_{\max}\) measures the maximum cohesive stress along the ideal opening path and provides an atomistic measure of resistance to the onset of local separation. Segregation energies identify preferred fission-product sites, and decorated GBs are then evaluated using the same decohesion analysis. Retention maps compare the strength penalty with the segregation driving force, and bonding analysis identifies the local electronic changes behind that comparison. The key question is whether fission products that segregate to internal interfaces also produce severe loss of GB strength. Details of the supercell construction, segregation-energy definition, decohesion procedure, and bonding analysis are given in the Methods.

\begin{figure}[t]
\centering
\includegraphics[width=1.0\textwidth]{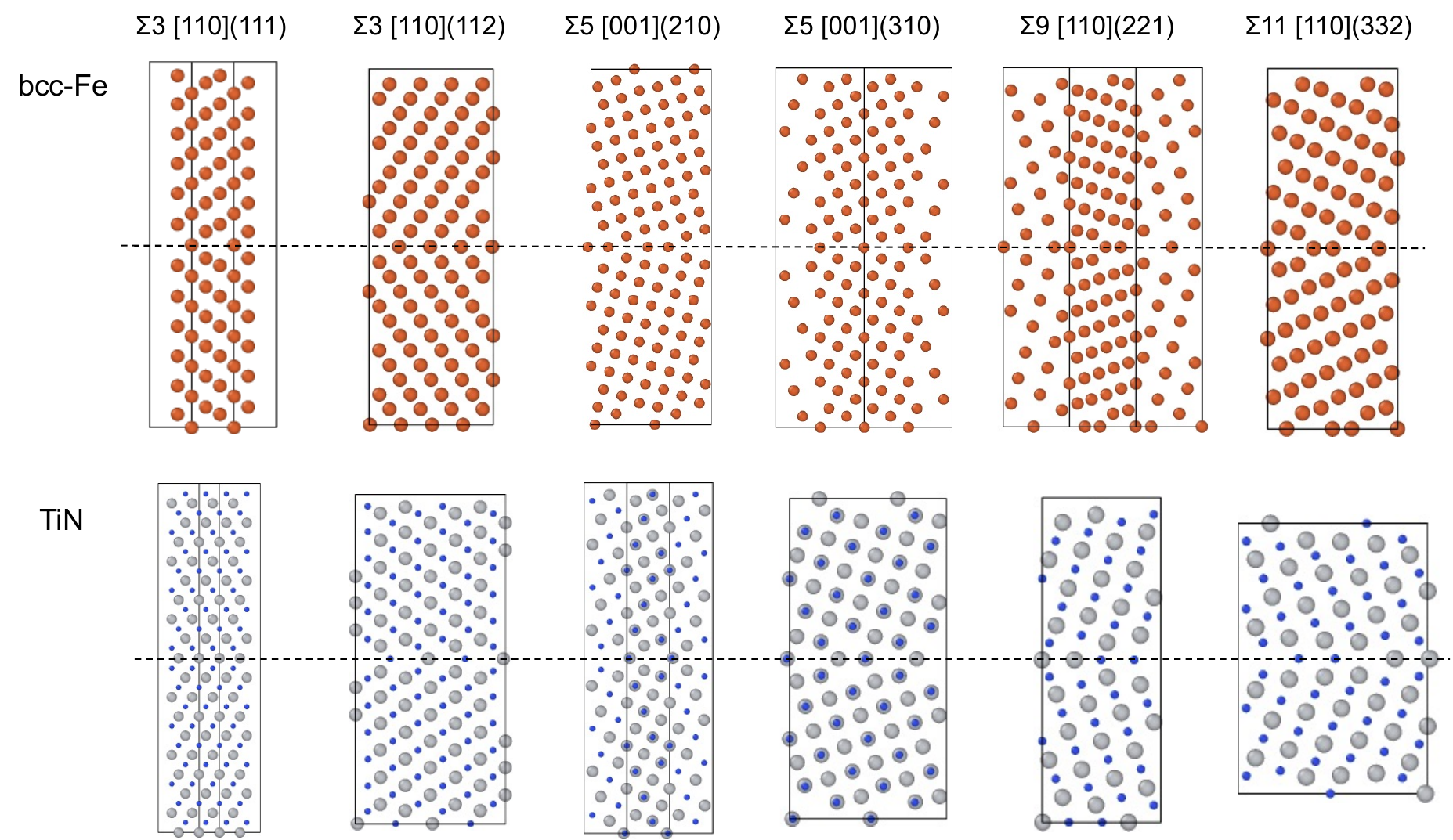}
\caption{Relaxed grain-boundary supercells obtained from DFT for bcc Fe and TiN, shown with periodic repeats for visualization. The top row shows bcc Fe and the bottom row shows TiN for the \(\Sigma 3\,[110](111)\), \(\Sigma 3\,[110](112)\), \(\Sigma 5\,[001](210)\), \(\Sigma 5\,[001](310)\), \(\Sigma 9\,[110](221)\), and \(\Sigma 11\,[110](332)\) CSL boundaries, from left to right. Orange atoms are Fe, gray atoms are Ti, and blue atoms are N.}
\label{fig:dft-gb-configuration}
\end{figure}

Figure~\ref{fig:dft-gb-configuration} shows the six CSL GB structures examined in both bcc Fe and TiN: \(\Sigma 3\,[110](111)\), \(\Sigma 3\,[110](112)\), \(\Sigma 5\,[001](210)\), \(\Sigma 5\,[001](310)\), \(\Sigma 9\,[110](221)\), and \(\Sigma 11\,[110](332)\).

Unless otherwise stated, a decorated GB denotes the best-site segregation configuration, defined as the configuration with the lowest \(E_{\mathrm{seg}}\) for a given material, GB, and fission product. The reported \(W_{\mathrm{sep}}\) values are direct DFT separation energies rather than asymptotes from the fitted cohesive-law form, whereas \(\sigma_{\max}\) is obtained from the fitted traction--separation curve. Because each decorated supercell contains one Nd or Ce atom, the results represent a dilute measure of fission-product effects on GB strength at specific sites.

\subsection{Clean GBs}
Calculation results for the clean bcc Fe and TiN GBs establish the reference states needed to interpret the fission-product-decorated boundaries. For Fe, the calculated \(\gamma_{\mathrm{GB}}\) values vary from \(0.44~\mathrm{J\,m^{-2}}\) for \(\Sigma 3\,[110](112)\) to \(1.70~\mathrm{J\,m^{-2}}\) for \(\Sigma 5\,[001](210)\), as summarized in Table~\ref{tab:clean-fe-baselines}. The direct DFT \(W_{\mathrm{sep}}\) values vary over a narrower interval, from \(3.63\) to \(4.84~\mathrm{J\,m^{-2}}\), and the corresponding peak tractions range from \(25.3\) to \(33.6~\mathrm{GPa}\). The \(\Sigma 3\,[110](112)\) boundary has the lowest \(\gamma_{\mathrm{GB}}\) and the largest \(W_{\mathrm{sep}}\) and \(\sigma_{\max}\), showing that the most thermodynamically stable Fe boundary in this set is not the most susceptible to ideal cleavage.

The clean Fe values also benchmark the computational approach. Wang et al. reported DFT GB energies for the same six bcc Fe CSL boundaries after sampling rigid-body translations~\cite{wang2018grain}, and Mai et al. reported GB energies and cohesive properties for the same set in a high-throughput first-principles study~\cite{mai2025high,mai2022segregation}. For these six GBs, the present \(\gamma_{\mathrm{GB}}\) values differ from the Wang et al. and Mai et al. values by at most \(0.06\) and \(0.12~\mathrm{J\,m^{-2}}\), respectively. For the four [110] boundaries also reported in the 2022 study~\cite{mai2022segregation}, the present direct \(W_{\mathrm{sep}}\) values differ from the relaxed \(W_{\mathrm{sep}}\) values of Mai et al. by at most \(0.20~\mathrm{J\,m^{-2}}\). The Mai et al. rigid-grain separation (RGS) \(W_{\mathrm{sep}}^{\mathrm{RGS}}\) values are listed separately because they correspond to separation without cleaved-surface relaxation. This agreement confirms that the clean-boundary reference states are consistent with previous first-principles Fe GB calculations.

\begin{table}[t]
\centering
\caption{Clean Fe GB excess energy, direct DFT work of separation, and peak
traction compared with corresponding clean bcc Fe reference values. Wang
et al.~\cite{wang2018grain} reported DFT GB energies. The work-of-separation
values in Mai et al.~\cite{mai2025high} correspond to rigid-grain separation
(RGS), whereas the values in Mai et al.~\cite{mai2022segregation} include
relaxed cleaved surfaces.}
\label{tab:clean-fe-baselines}
\tablefontsize
\setlength{\tabcolsep}{3pt}
\begin{tabular}{lccccc}
\toprule
GB &
\multicolumn{2}{c}{\(\gamma_{\mathrm{GB}}\)}
& \multicolumn{2}{c}{\(W_{\mathrm{sep}}\)}
& \(\sigma_{\max}\) \\
\cmidrule(lr){2-3}\cmidrule(lr){4-5}\cmidrule(lr){6-6}
& This work & Literature
& This work & Literature
& This work \\
& \multicolumn{2}{c}{\((\mathrm{J\,m^{-2}})\)}
& \multicolumn{2}{c}{\((\mathrm{J\,m^{-2}})\)}
& \multicolumn{1}{c}{\((\mathrm{GPa})\)} \\
\midrule
\(\Sigma 3\,[110](111)\) & 1.58 & 1.57~\cite{wang2018grain}, 1.58~\cite{mai2025high} & 4.01 & 3.81~\cite{mai2022segregation}, 4.19~\cite{mai2025high} & 29.66 \\
\(\Sigma 3\,[110](112)\) & 0.44 & 0.45~\cite{wang2018grain}, 0.45~\cite{mai2025high} & 4.84 & 4.72~\cite{mai2022segregation}, 4.88~\cite{mai2025high} & 33.55 \\
\(\Sigma 5\,[001](210)\) & 1.70 & 1.64~\cite{wang2018grain}, 1.62~\cite{mai2025high} & 3.65 & 3.96~\cite{mai2025high} & 25.29 \\
\(\Sigma 5\,[001](310)\) & 1.57 & 1.57~\cite{wang2018grain}, 1.69~\cite{mai2025high} & 3.63 & 3.73~\cite{mai2025high} & 26.37 \\
\(\Sigma 9\,[110](221)\) & 1.64 & 1.66~\cite{wang2018grain}, 1.75~\cite{mai2025high} & 3.76 & 3.60~\cite{mai2022segregation}, 4.22~\cite{mai2025high} & 25.40 \\
\(\Sigma 11\,[110](332)\) & 1.42 & 1.47~\cite{wang2018grain}, 1.45~\cite{mai2025high} & 3.93 & 3.83~\cite{mai2022segregation}, 4.27~\cite{mai2025high} & 25.86 \\
\bottomrule
\end{tabular}
\end{table}

The clean TiN GBs display a wider range of strength response than the clean Fe GBs, as shown in Table~\ref{tab:clean-tin-baselines}. The calculated \(\gamma_{\mathrm{GB}}\) values range from \(0.34~\mathrm{J\,m^{-2}}\) for \(\Sigma 3\,[110](111)\) to \(2.56~\mathrm{J\,m^{-2}}\) for \(\Sigma 11\,[110](332)\). The clean \(W_{\mathrm{sep}}\) values span \(2.38\) to \(8.88~\mathrm{J\,m^{-2}}\), and the corresponding \(\sigma_{\max}\) values span \(22.1\) to \(74.1~\mathrm{GPa}\). The \(\Sigma 3\,[110](111)\) TiN boundary is the strongest boundary in the present dataset and also has the lowest TiN GB excess energy, whereas the two sampled \(\Sigma 5\) TiN boundaries have lower separation works and peak tractions than several Fe boundaries. Triestram and Polfus reported DFT excess energies for open and compact \(\Sigma 5\,[001](310)\) and \(\Sigma 5\,[001](210)\) TiN GBs~\cite{triestram2024hydride}. The present \(\Sigma 5\) TiN cells correspond to the compact variants based on stoichiometry and atom count, so the compact literature values are listed for comparison.

\begin{table}[t]
\centering
\caption{Clean TiN GB excess energy, direct DFT work of separation, and peak
traction. Literature values are DFT GB excess energies from Triestram and
Polfus~\cite{triestram2024hydride} for the compact \(\Sigma 5\) structures.}
\label{tab:clean-tin-baselines}
\tablefontsize
\setlength{\tabcolsep}{4pt}
\begin{tabular}{lcccc}
\toprule
GB &
\multicolumn{2}{c}{\(\gamma_{\mathrm{GB}}\)} &
\(W_{\mathrm{sep}}\) &
\(\sigma_{\max}\) \\
\cmidrule(lr){2-3}\cmidrule(lr){4-4}\cmidrule(lr){5-5}
& This work & Literature
& This work
& This work \\
& \multicolumn{2}{c}{\((\mathrm{J\,m^{-2}})\)}
& \multicolumn{1}{c}{\((\mathrm{J\,m^{-2}})\)}
& \multicolumn{1}{c}{\((\mathrm{GPa})\)} \\
\midrule
\(\Sigma 3\,[110](111)\) & 0.34 & -- & 8.88 & 74.15 \\
\(\Sigma 3\,[110](112)\) & 2.09 & -- & 4.30 & 36.13 \\
\(\Sigma 5\,[001](210)\) & 1.82 & 1.90~\cite{triestram2024hydride} & 2.74 & 22.74 \\
\(\Sigma 5\,[001](310)\) & 1.89 & 1.79~\cite{triestram2024hydride} & 2.38 & 22.07 \\
\(\Sigma 9\,[110](221)\) & 2.45 & -- & 4.63 & 32.04 \\
\(\Sigma 11\,[110](332)\) & 2.56 & -- & 5.11 & 43.70 \\
\bottomrule
\end{tabular}
\end{table}

\begin{figure}[t]
\centering
\includegraphics[width=0.96\textwidth]{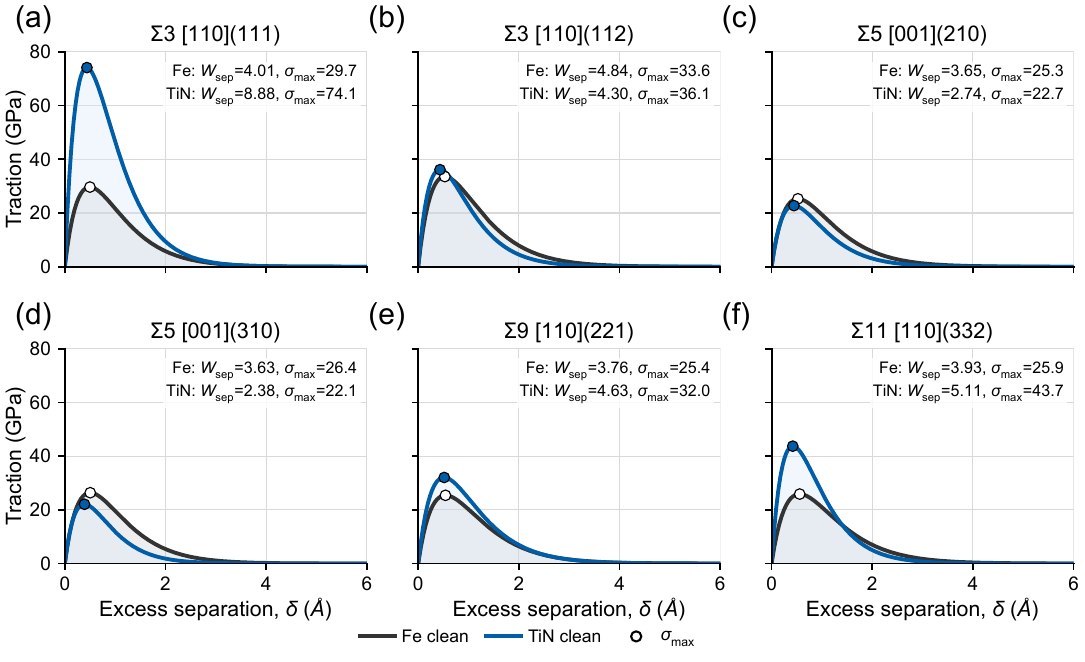}
\caption{Clean traction--separation curves for bcc Fe and TiN GBs with the same CSL geometries.
(a)--(f) correspond to \(\Sigma 3\,[110](111)\), \(\Sigma 3\,[110](112)\), \(\Sigma 5\,[001](210)\), \(\Sigma 5\,[001](310)\), \(\Sigma 9\,[110](221)\), and \(\Sigma 11\,[110](332)\), respectively. Peak markers indicate \(\sigma_{\max}\), and the shaded area illustrates the separation work associated with the fitted cohesive law. In-figure \(W_{\mathrm{sep}}\) and \(\sigma_{\max}\) values are in \(\mathrm{J\,m^{-2}}\) and GPa, respectively.}
\label{fig:clean-fe-tin-traction}
\end{figure}

The same clean strength responses are shown as fitted traction--separation curves in Fig.~\ref{fig:clean-fe-tin-traction}, where (a)--(f) compare the same CSL geometries in bcc Fe and TiN. The maximum of each curve defines the interfacial strength, \(\sigma_{\max}\), while the area under the fitted curve illustrates the separation work associated with the cohesive law, \(W_{\mathrm{sep}}=\int T(\delta)\,\mathrm{d}\delta\). In the plotted units, \(1~\mathrm{GPa}\,\mathring{\mathrm{A}}=0.1~\mathrm{J\,m^{-2}}\). The \(W_{\mathrm{sep}}\) values reported in Tables~\ref{tab:clean-fe-baselines} and~\ref{tab:clean-tin-baselines} are direct DFT total-energy differences, and the continuous curves give the smooth derivative used to extract \(\sigma_{\max}\).

The clean-boundary results set the Fe and TiN baselines needed to evaluate fission-product effects. They should not be read as a simple ranking of Fe and TiN for each CSL geometry. As shown in Fig.~\ref{fig:clean-fe-tin-traction}~(a), the \(\Sigma 3\,[110](111)\) TiN boundary has both larger \(W_{\mathrm{sep}}\) and larger \(\sigma_{\max}\) than its bcc Fe counterpart, whereas Fig.~\ref{fig:clean-fe-tin-traction}~(c) and (d) show that the two clean \(\Sigma 5\) TiN boundaries have lower separation works and peak tractions than several Fe boundaries. TiN is therefore not intrinsically stronger than bcc Fe for every idealized interface. The decorated-boundary calculations below use these clean states as the reference.

\subsection{Segregation}

Calculation results show that Nd and Ce are thermodynamically favored at the best-site GB configurations in Fe and TiN. In Fe, the best-site \(E_{\mathrm{seg}}\) values for Nd range from \(-2.80\) to \(-0.24~\mathrm{eV}\), with a median value of \(-1.56~\mathrm{eV}\), while the corresponding Ce values range from \(-2.44\) to \(-0.26~\mathrm{eV}\), with a median value of \(-1.51~\mathrm{eV}\). In TiN, the best-site Nd values range from \(-4.30\) to \(-0.12~\mathrm{eV}\), with a median value of \(-3.15~\mathrm{eV}\), and the best-site Ce values range from \(-4.78\) to \(-0.07~\mathrm{eV}\), with a median value of \(-3.48~\mathrm{eV}\).

The per-boundary values in Table~\ref{tab:best-site-segregation} show that the more favorable median segregation energies in TiN are not confined to one boundary. Several TiN boundaries are stronger thermodynamic trapping sites than the corresponding Fe boundaries, although the \(\Sigma 3\,[110](111)\) TiN boundary is an exception with only weakly favorable Nd and Ce segregation. This result indicates that Nd and Ce can be stabilized at TiN interfaces. We next compare whether this stabilization produces the same strength loss in Fe and TiN.

\begin{table}[t]
\centering
\caption{Best-site segregation energies for Nd and Ce at the sampled Fe and TiN GBs. Negative values indicate favorable segregation to the GB.}
\label{tab:best-site-segregation}
\setlength{\tabcolsep}{5pt}
\tablefontsize
\begin{tabular}{lcccc}
\toprule
GB & Fe--Nd & Fe--Ce & TiN--Nd & TiN--Ce \\
& \multicolumn{4}{c}{\(E_{\mathrm{seg}}\) \((\mathrm{eV})\)} \\
\midrule
\(\Sigma 3\,[110](111)\) & -1.55 & -1.48 & -0.12 & -0.07 \\
\(\Sigma 3\,[110](112)\) & -0.24 & -0.26 & -2.53 & -3.00 \\
\(\Sigma 5\,[001](210)\) & -1.99 & -2.06 & -3.06 & -3.25 \\
\(\Sigma 5\,[001](310)\) & -1.16 & -1.16 & -4.00 & -3.98 \\
\(\Sigma 9\,[110](221)\) & -2.80 & -2.44 & -4.30 & -4.78 \\
\(\Sigma 11\,[110](332)\) & -1.57 & -1.54 & -3.24 & -3.70 \\
\bottomrule
\end{tabular}
\end{table}

\subsection{Decohesion response of Fe}

\begin{figure}[t]
\centering
\includegraphics[width=1.00\textwidth]{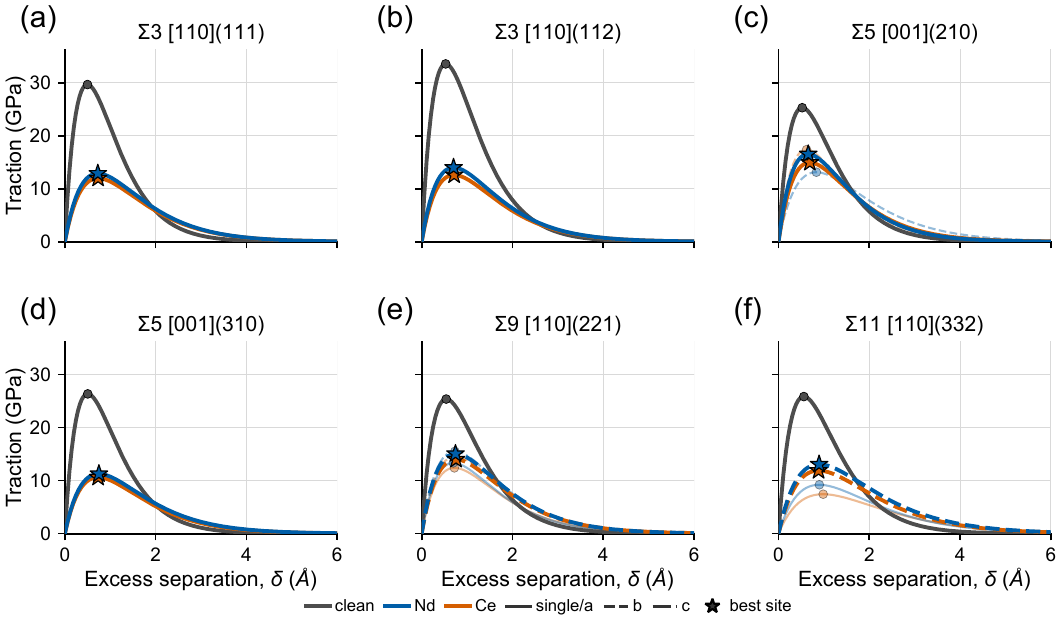}
\caption{Fe traction--separation curves from the \textit{ab initio} tensile
calculations for clean GBs and Nd/Ce-decorated site variants. (a)--(f) correspond to \(\Sigma 3\,[110](111)\), \(\Sigma 3\,[110](112)\), \(\Sigma 5\,[001](210)\), \(\Sigma 5\,[001](310)\), \(\Sigma 9\,[110](221)\), and \(\Sigma 11\,[110](332)\), respectively. Stars identify the best-site configurations used for the summary metrics, and the maximum traction defines \(\sigma_{\max}\).}
\label{fig:fe-traction-curves}
\end{figure}

The Fe calculations show the clearest embrittling response. As shown in Fig.~\ref{fig:fe-traction-curves}~(a)--(f), Nd and Ce lower the traction over much of the decohesion path and reduce the maximum traction for every Fe GB. Across the six GBs, the mean changes relative to the corresponding clean GBs are \(\Delta W_{\mathrm{sep}}=-1.21~\mathrm{J\,m^{-2}}\) and \(\Delta\sigma_{\max}=-13.9~\mathrm{GPa}\) for Nd, and \(\Delta W_{\mathrm{sep}}=-1.42~\mathrm{J\,m^{-2}}\) and \(\Delta\sigma_{\max}=-15.0~\mathrm{GPa}\) for Ce. The peak traction is more sensitive to fission-product segregation than the integrated work of separation. No best-site Fe configuration shows an increase in either \(W_{\mathrm{sep}}\) or \(\sigma_{\max}\). The corresponding absolute best-site decorated \(W_{\mathrm{sep}}\) and \(\sigma_{\max}\) values for Fe and TiN are collected in Table~\ref{tab:decorated-cohesion}.

\begin{figure}[t]
\centering
\includegraphics[width=1\textwidth]{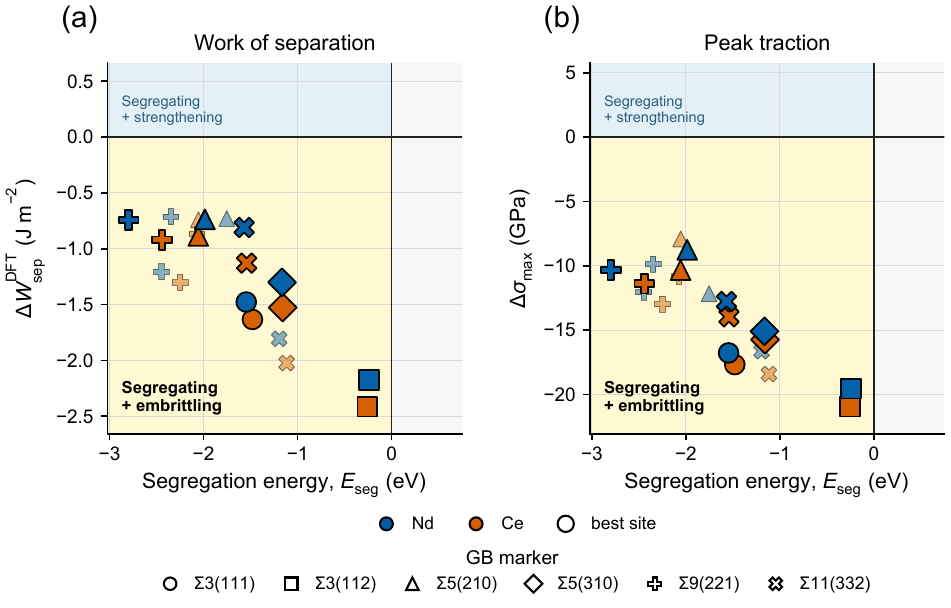}
\caption{Segregation--embrittlement map for Nd and Ce in Fe. Negative
\(E_{\mathrm{seg}}\) denotes favorable segregation. (a) and (b) show changes in \(W_{\mathrm{sep}}\) and \(\sigma_{\max}\), respectively. Negative changes in
\(W_{\mathrm{sep}}\) or \(\sigma_{\max}\) denote strength loss relative
to the clean GB, while positive changes denote preserved or enhanced strength. Larger, opaque symbols denote best-site configurations, while lighter symbols show additional sampled site variants.}
\label{fig:fe-segregation-embrittlement}
\end{figure}

The most severe Fe case is Ce at the \(\Sigma 3\,[110](112)\) boundary. This configuration reduces \(W_{\mathrm{sep}}\) by \(2.42~\mathrm{J\,m^{-2}}\) and \(\sigma_{\max}\) by \(21.0~\mathrm{GPa}\), although \(E_{\mathrm{seg}}\) is only \(-0.26~\mathrm{eV}\). The degradation affects the full decohesion trajectory rather than only the final separation energy. The segregation--embrittlement maps in Fig.~\ref{fig:fe-segregation-embrittlement}~(a) and (b) summarize the Fe response in \(W_{\mathrm{sep}}\) and \(\sigma_{\max}\), respectively. Every best-site Nd and Ce configuration combines favorable segregation with reductions in both metrics. Thus, the sampled Fe boundaries consistently show embrittlement after Nd/Ce segregation.

\subsection{Decohesion response of TiN}

\begin{figure}[t]
\centering
\includegraphics[width=1\textwidth]{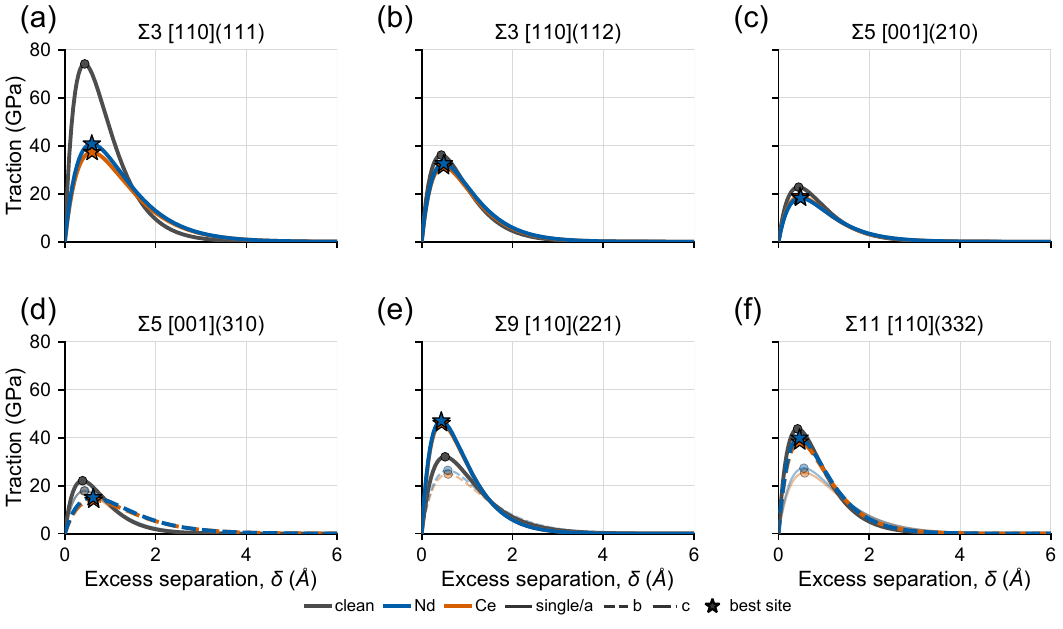}
\caption{TiN traction--separation curves from the \textit{ab initio} tensile
calculations for clean GBs and Nd/Ce-decorated site variants. (a)--(f) correspond to \(\Sigma 3\,[110](111)\), \(\Sigma 3\,[110](112)\), \(\Sigma 5\,[001](210)\), \(\Sigma 5\,[001](310)\), \(\Sigma 9\,[110](221)\), and \(\Sigma 11\,[110](332)\), respectively. Stars identify the best-site configurations used for the summary metrics.}
\label{fig:tin-traction-curves}
\end{figure}

The TiN calculations show a different response. Fig.~\ref{fig:tin-traction-curves}~(a)--(f) shows no uniform downward shift in the traction--separation curves after Nd or Ce segregation. Several decorated TiN boundaries maintain a traction response close to the clean curve, and some have higher separation work or peak traction than the corresponding clean GB. Across the six GBs, the mean changes relative to the corresponding clean GBs are \(\Delta W_{\mathrm{sep}}=-0.23~\mathrm{J\,m^{-2}}\) and \(\Delta\sigma_{\max}=-6.2~\mathrm{GPa}\) for Nd, and \(\Delta W_{\mathrm{sep}}=-0.39~\mathrm{J\,m^{-2}}\) and \(\Delta\sigma_{\max}=-7.5~\mathrm{GPa}\) for Ce. Three of the six best-site Nd configurations and two of the six best-site Ce configurations increase \(W_{\mathrm{sep}}\) relative to the clean TiN GB, while one best-site configuration for each fission product increases \(\sigma_{\max}\). This response contrasts with Fe, where all decorated GBs are weakened in both metrics.

The most degraded TiN case is the \(\Sigma 3\,[110](111)\) boundary shown in Fig.~\ref{fig:tin-traction-curves}~(a). Ce reduces \(W_{\mathrm{sep}}\) by \(2.73~\mathrm{J\,m^{-2}}\) and \(\sigma_{\max}\) by \(36.8~\mathrm{GPa}\), while Nd reduces them by \(2.24~\mathrm{J\,m^{-2}}\) and \(33.4~\mathrm{GPa}\). It is worth noting that this boundary still maintains substantial GB strength after decoration, with \(W_{\mathrm{sep}}\) values of \(6.16~\mathrm{J\,m^{-2}}\) for Ce and \(6.65~\mathrm{J\,m^{-2}}\) for Nd and corresponding peak tractions of \(37.3\) and \(40.8~\mathrm{GPa}\) (Table~\ref{tab:decorated-cohesion}). The same boundary also has only weakly favorable segregation, with best-site \(E_{\mathrm{seg}}\) values of \(-0.07~\mathrm{eV}\) for Ce and \(-0.12~\mathrm{eV}\) for Nd. These magnitudes are comparable to \(k_{\mathrm{B}}T\) at \(600^{\circ}\mathrm{C}\) (\(\sim 0.075~\mathrm{eV}\)). Therefore, the large fractional decrease relative to the unusually strong clean \(\Sigma 3\,[110](111)\) TiN boundary should not be interpreted as Fe-like embrittlement.

\begin{figure}[t]
\centering
\includegraphics[width=1\textwidth]{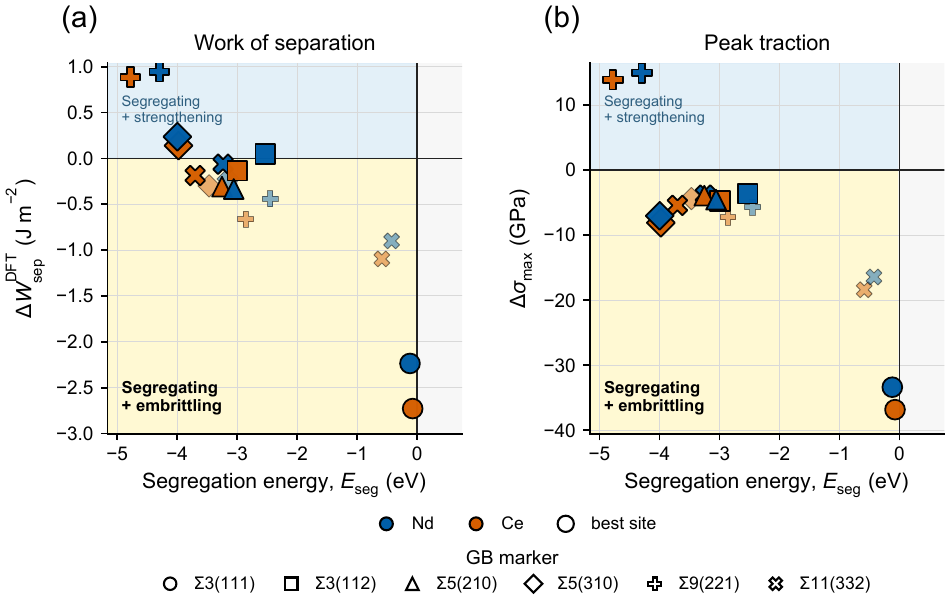}
\caption{Segregation--embrittlement map for Nd and Ce in TiN. Negative
\(E_{\mathrm{seg}}\) denotes favorable segregation. (a) and (b) show changes in \(W_{\mathrm{sep}}\) and \(\sigma_{\max}\), respectively. Negative changes denote strength loss relative to the clean GB, while positive changes denote preserved or enhanced strength. Larger, opaque symbols denote best-site configurations, while lighter symbols show additional sampled site variants.}
\label{fig:tin-segregation-embrittlement}
\end{figure}

The segregation--embrittlement maps in Fig.~\ref{fig:tin-segregation-embrittlement}~(a) and (b) show the same contrast for \(W_{\mathrm{sep}}\) and \(\sigma_{\max}\). Unlike Fe, where the best-site configurations lie well inside the favorable-segregation/strength-loss region in Fig.~\ref{fig:fe-segregation-embrittlement}~(a) and (b), the TiN points are distributed close to the edge of this region or extend into preserved and enhanced strength. Several TiN GBs with strongly negative \(E_{\mathrm{seg}}\) maintain or increase these strength metrics. This result shows that the largest mechanical loss in TiN does not coincide with the strongest segregation driving force.

To complement the changes in \(W_{\mathrm{sep}}\) and \(\sigma_{\max}\), Table~\ref{tab:decorated-cohesion} lists the absolute best-site decorated values for all sampled GBs. Decorated TiN has the larger peak traction for every sampled GB and fission product. It also has the larger \(W_{\mathrm{sep}}\) for all corresponding Fe/TiN comparisons except \(\Sigma 5\,[001](210)\). These values provide the strength scale behind the normalized comparisons that follow.

\begin{table}[t]
\centering
\caption{Absolute best-site decorated strength metrics for Nd and Ce in Fe and TiN. \(W_{\mathrm{sep}}\) is in \(\mathrm{J\,m^{-2}}\), and \(\sigma_{\max}\) is in GPa.}
\label{tab:decorated-cohesion}
\setlength{\tabcolsep}{2.0pt}
\tablefontsize
\begin{tabular}{lcccccccc}
\toprule
GB &
\multicolumn{2}{c}{Fe--Nd} &
\multicolumn{2}{c}{Fe--Ce} &
\multicolumn{2}{c}{TiN--Nd} &
\multicolumn{2}{c}{TiN--Ce} \\
& \(W_{\mathrm{sep}}\) & \(\sigma_{\max}\)
& \(W_{\mathrm{sep}}\) & \(\sigma_{\max}\)
& \(W_{\mathrm{sep}}\) & \(\sigma_{\max}\)
& \(W_{\mathrm{sep}}\) & \(\sigma_{\max}\) \\
\midrule
\(\Sigma 3\,[110](111)\) & 2.53 & 12.9 & 2.38 & 12.0 & 6.65 & 40.8 & 6.16 & 37.3 \\
\(\Sigma 3\,[110](112)\) & 2.66 & 14.0 & 2.42 & 12.6 & 4.35 & 32.6 & 4.17 & 31.4 \\
\(\Sigma 5\,[001](210)\) & 2.91 & 16.5 & 2.76 & 14.9 & 2.41 & 18.2 & 2.43 & 18.8 \\
\(\Sigma 5\,[001](310)\) & 2.33 & 11.3 & 2.11 & 10.6 & 2.61 & 15.1 & 2.52 & 14.0 \\
\(\Sigma 9\,[110](221)\) & 3.02 & 15.1 & 2.84 & 14.0 & 5.58 & 47.0 & 5.52 & 46.0 \\
\(\Sigma 11\,[110](332)\) & 3.12 & 13.1 & 2.80 & 11.9 & 5.05 & 39.9 & 4.92 & 38.3 \\
\bottomrule
\end{tabular}
\end{table}

\subsection{Strength retention and segregation tolerance}
The results separate the thermodynamic preference for fission-product segregation from its mechanical consequence. \(E_{\mathrm{seg}}\) gives the driving force for Nd or Ce to occupy a GB site, while the changes in \(W_{\mathrm{sep}}\) and \(\sigma_{\max}\) describe the strength penalty after that atom is incorporated. For coating degradation, a GB can trap fission products without showing a large loss in ideal cohesive strength. In the Rice--Wang framework, the embrittling effect is governed by the relative stabilization of the intact GB and the separated fracture surfaces \cite{rice1989embrittlement}. From the definition \(W_{\mathrm{sep}}=(E_{\mathrm{sep}}-E_{\mathrm{GB}})/A\), segregation reduces \(W_{\mathrm{sep}}\) when it lowers \(E_{\mathrm{sep}}\) more than \(E_{\mathrm{GB}}\). A strongly segregated fission product therefore does not necessarily produce a large strength loss if the separated surfaces are not stabilized by a comparable or larger amount.

\begin{figure}[!htbp]
\centering
\includegraphics[width=0.40\textwidth]{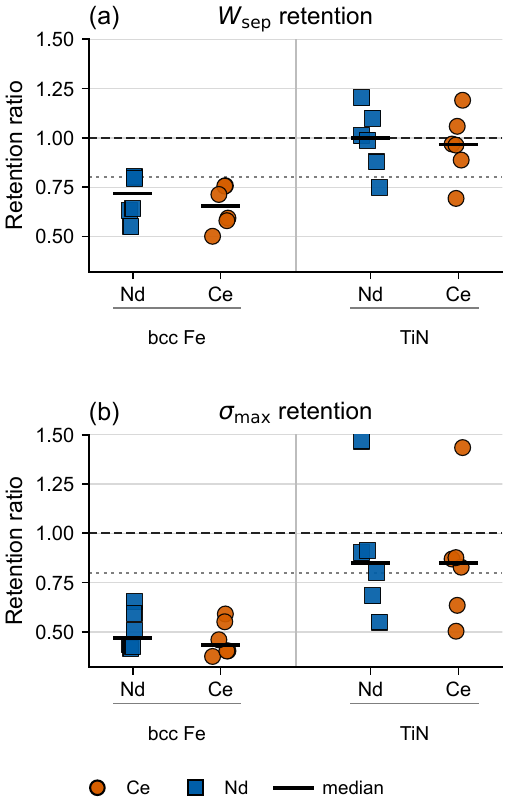}
\caption{Best-site retention envelope for Nd and Ce in Fe and TiN. (a) and (b) show retention ratios for \(W_{\mathrm{sep}}\) and \(\sigma_{\max}\), respectively. Each point represents one sampled GB for a given material and fission product. The dashed line corresponds to unchanged strength, and black bars indicate medians.}
\label{fig:retention-envelope}
\end{figure}

To compare the mechanical response after segregation on a common scale, the retention of a strength-related quantity \(Q\) is defined as
\begin{equation}
R_Q =
\frac{Q^{\mathrm{decorated}}}{Q^{\mathrm{clean}}},
\label{eq:retention}
\end{equation}
where \(Q\) represents either \(W_{\mathrm{sep}}\) or \(\sigma_{\max}\). Thus, \(W_{\mathrm{sep}}\) retention and \(\sigma_{\max}\) retention give the fractions of the clean separation work and peak traction preserved after segregation. Using Eq.~\eqref{eq:retention}, Fig.~\ref{fig:retention-envelope}~(a) and Table~\ref{tab:retention-summary} show that Fe preserves only 72\% of its median clean \(W_{\mathrm{sep}}\) after Nd segregation and 65\% after Ce segregation. Fig.~\ref{fig:retention-envelope}~(b) shows the larger peak-traction penalty in Fe, with retentions of 47\% and 43\% for Nd and Ce. TiN preserves 100\% and 97\% of its median clean \(W_{\mathrm{sep}}\) for Nd and Ce, respectively, and preserves 85\% of its median clean \(\sigma_{\max}\) for both fission products. The absolute values in Table~\ref{tab:decorated-cohesion} give the corresponding strength scale. These values show the different mechanical penalty associated with segregation in the two materials.

\begin{table}[!htbp]
\centering
\caption{Median best-site retention metrics across the six sampled GBs.}
\label{tab:retention-summary}
\tablefontsize
\begin{tabular}{llcc}
\toprule
Material & Fission product & \(W_{\mathrm{sep}}\) retention & \(\sigma_{\max}\) retention \\
\midrule
Fe & Nd & 0.72 & 0.47 \\
Fe & Ce & 0.65 & 0.43 \\
TiN & Nd & 1.00 & 0.85 \\
TiN & Ce & 0.97 & 0.85 \\
\bottomrule
\end{tabular}
\end{table}

\begin{figure}[!htbp]
\centering
\includegraphics[width=0.76\textwidth]{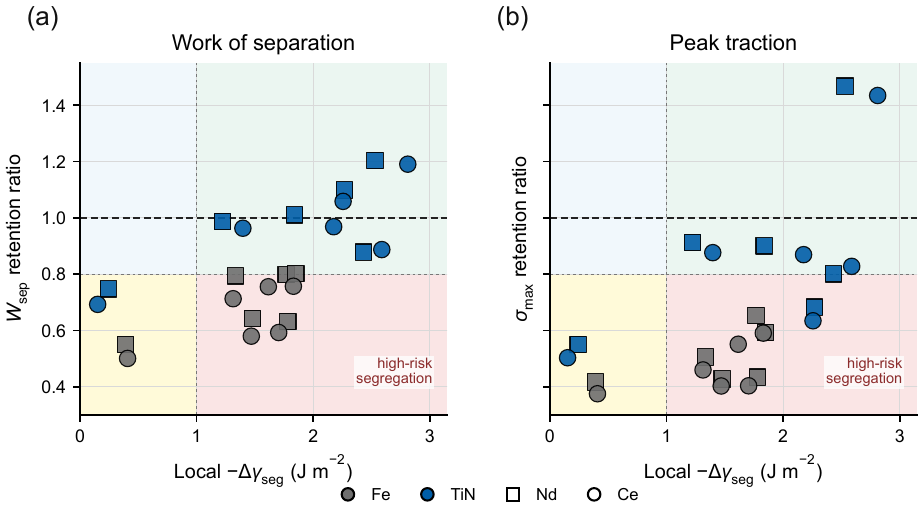}
\caption{Area-normalized segregation tolerance map. (a) and (b) show retention ratios for \(W_{\mathrm{sep}}\) and \(\sigma_{\max}\), respectively. Because favorable segregation gives negative \(\Delta\gamma_{\mathrm{seg}}^{\mathrm{local}}\), the horizontal axis uses the positive reduction \(-\Delta\gamma_{\mathrm{seg}}^{\mathrm{local}}\), where \(\Delta\gamma_{\mathrm{seg}}^{\mathrm{local}}=E_{\mathrm{seg}}/A\). Horizontal dashed lines indicate 80\% and 100\% retention, and the vertical dashed line corresponds to \(1~\mathrm{J\,m^{-2}}\). The dashed lines are visual guides for comparing the present dataset, not universal embrittlement criteria.}
\label{fig:delta-gamma-tolerance}
\end{figure}

Area normalization gives the same comparison in terms of segregation energy per interface area. The local area-normalized segregation contribution is \(\Delta\gamma_{\mathrm{seg}}^{\mathrm{local}}=E_{\mathrm{seg}}/A\), where \(A\) is the GB area. With this sign convention, favorable segregation gives negative \(\Delta\gamma_{\mathrm{seg}}^{\mathrm{local}}\), so the maps use the positive quantity \(-\Delta\gamma_{\mathrm{seg}}^{\mathrm{local}}\). The median local segregation-energy reductions, \(-\Delta\gamma_{\mathrm{seg}}^{\mathrm{local}}\), in TiN are \(2.05~\mathrm{J\,m^{-2}}\) for Nd and \(2.22~\mathrm{J\,m^{-2}}\) for Ce, compared with \(1.62~\mathrm{J\,m^{-2}}\) and \(1.54~\mathrm{J\,m^{-2}}\) in Fe. TiN shows stronger Nd/Ce trapping per unit area than Fe, but most TiN points remain in the high-retention regions of Fig.~\ref{fig:delta-gamma-tolerance}~(a) and (b). Fe occupies lower-retention regions despite having a smaller median segregation driving force per unit area. A high-risk interfacial degradation case therefore requires favorable trapping together with a large reduction in \(W_{\mathrm{sep}}\) or \(\sigma_{\max}\). The horizontal 80\% retention lines in Fig.~\ref{fig:delta-gamma-tolerance}~(a) and (b) are visual guides corresponding to a 20\% loss of strength response, not universal embrittlement criteria.

\FloatBarrier
\subsection{Local bonding changes after fission-product segregation}
The retention contrast reflects different local bonding responses in Fe and TiN. In a local-orbital description, crystal-orbital overlap population (COOP) measures whether occupied electronic states have bonding or antibonding overlap character between selected atom pairs, whereas crystal-orbital Hamilton population (COHP) partitions the band-structure energy into pairwise orbital interactions. In the convention used here, positive \(-\mathrm{COHP}\) indicates bonding contributions, and a larger integrated \(-\mathrm{COHP}\) over occupied states corresponds to a larger bonding-energy contribution for the selected bond group. COHP is therefore closer to the energy balance that controls \(W_{\mathrm{sep}}\), while COOP provides an independent sign check on whether occupied states contain antibonding overlap population.

These quantities help connect the bonding calculations to environmentally assisted intergranular fracture and corrosion-cracking susceptibility at the atomistic scale. Fission-product-assisted cracking requires chemical redistribution and a mechanical separation path. The present calculations do not model electrochemical reactions or crack advance, but they test whether Nd/Ce segregation leaves the GB with weaker occupied bonding states and lower resistance to ideal opening. If a segregant replaces a strong matrix bond with a weak or partially antibonding interaction, the interfacial energy balance can shift toward decohesion. If the surrounding bonding network is preserved, segregation can occur without the same loss of GB strength.

\begin{figure}[!htbp]
\centering
\includegraphics[width=0.96\textwidth]{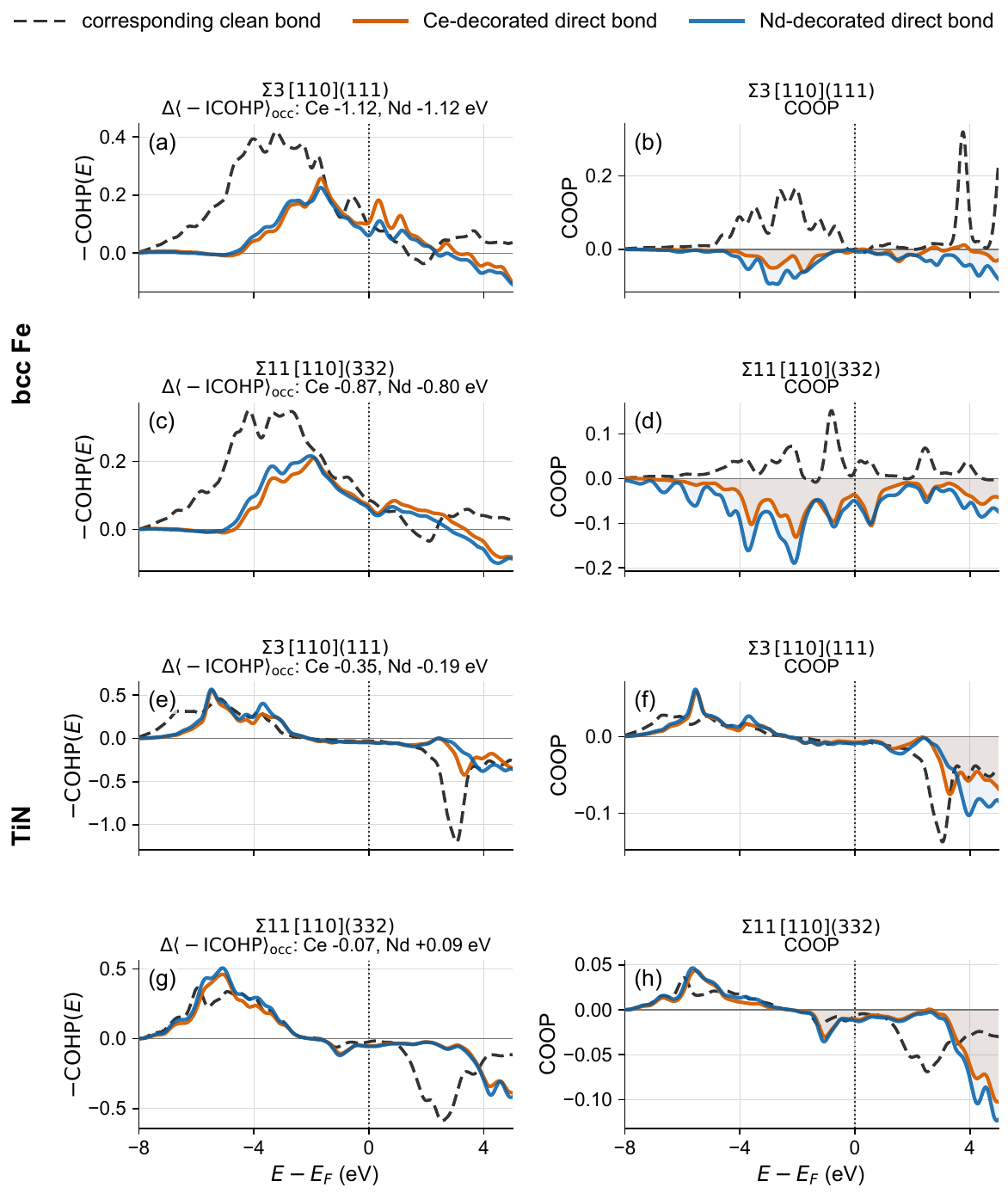}
\caption{Energy-resolved LOBSTER bonding comparison for corresponding Nd- and Ce-decorated Fe and TiN GBs. (a)--(d) show bcc Fe and (e)--(h) show TiN. The columns show \(-\mathrm{COHP}(E)\) and COOP for \(\Sigma 3\,[110](111)\) and \(\Sigma 11\,[110](332)\). Black dashed curves are the corresponding clean bonds, Fe--Fe in Fe and Ti--N in TiN. Solid orange and blue curves are the decorated direct bonds, X--Fe in Fe and X--N in TiN, where X is Ce or Nd. Orange and blue correspond to Ce and Nd, respectively. Vertical dotted lines indicate the Fermi level. The curves are plotted from \(-8\) to \(5~\mathrm{eV}\) for clarity, while the reported \(\Delta\langle-\mathrm{ICOHP}\rangle_{\mathrm{occ}}\) is calculated using the window-integrated COHP definition given in the Methods.}
\label{fig:corresponding-lobster-cohp-coop}
\end{figure}

\begin{figure}[!htbp]
\centering
\includegraphics[width=0.55\textwidth]{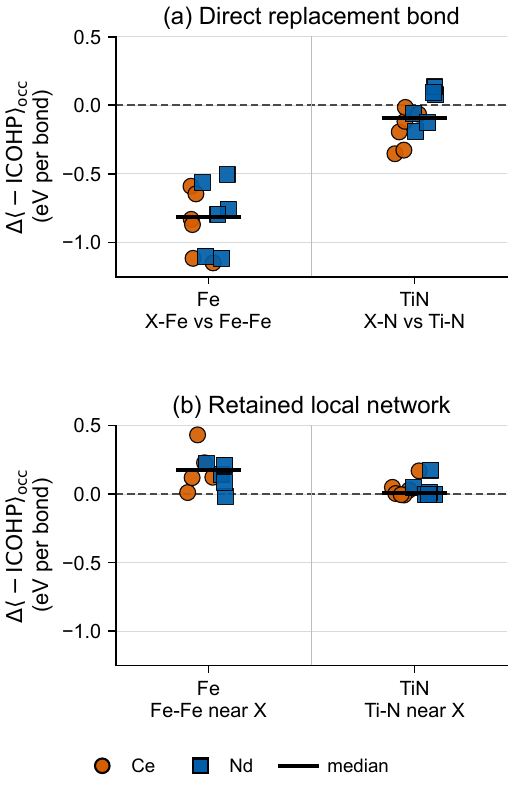}
\caption{LOBSTER bonding summary for all six sampled GBs using clean structures and best-site structures decorated with Nd or Ce. The plotted values are \(\Delta\langle-\mathrm{ICOHP}\rangle_{\mathrm{occ}}\), calculated from the occupied, window-integrated COHP values described in the Methods. (a) compares direct replacement bonds, X--Fe in Fe and X--N in TiN, against the corresponding clean bond group. (b) compares the preserved local network: nearby Fe--Fe bonds in Fe and Ti--N bonds in TiN that do not contain X. Circles denote Ce, squares denote Nd, and black bars indicate medians.}
\label{fig:lobster-bonding-summary}
\end{figure}

The bonding analysis also uses the integrated crystal-orbital bond index (ICOBI) as a complementary measure of bond order, but the COHP metric is expected to track strength loss more directly because it is energy-based. The Local Orbital Basis Suite Toward Electronic-Structure Reconstruction (LOBSTER) analysis compared two bond groups: the direct bonds formed by Nd/Ce after substitution and the nearby Fe--Fe or Ti--N network that remains around the segregant. Fig.~\ref{fig:corresponding-lobster-cohp-coop}~(a)--(h) compares energy-resolved COHP and COOP curves for best-site Ce- and Nd-decorated \(\Sigma 3\,[110](111)\) and \(\Sigma 11\,[110](332)\) GBs in both Fe and TiN. These boundaries sample an ordered low-\(\Sigma\) interface and a higher-index interface. Fig.~\ref{fig:lobster-bonding-summary}~(a) and (b) summarizes the occupied, integrated COHP change, \(\Delta\langle-\mathrm{ICOHP}\rangle_{\mathrm{occ}}\), from the full LOBSTER set, including all six clean GBs and the corresponding best-site structures decorated with Nd or Ce.

The energy-resolved curves in Fig.~\ref{fig:corresponding-lobster-cohp-coop}~(a) and (b) show why Fe is sensitive to fission-product segregation. For the Fe \(\Sigma 3\,[110](111)\) boundary in Fig.~\ref{fig:corresponding-lobster-cohp-coop}~(a), both Ce--Fe and Nd--Fe direct bonds have much less occupied bonding intensity than the clean Fe--Fe bond at the same site, shown by the black dashed curve, with \(\Delta\langle-\mathrm{ICOHP}\rangle_{\mathrm{occ}}\approx -1.12~\mathrm{eV}\) per bond for both fission products. The corresponding COOP curves in Fig.~\ref{fig:corresponding-lobster-cohp-coop}~(b) change from mostly positive overlap population for the clean Fe--Fe reference, again shown by the black dashed curve, to weak or negative overlap population for X--Fe over much of the occupied energy range from \(-10~\mathrm{eV}\) to \(E_F\). This negative COOP contribution is consistent with significant occupied antibonding character in the direct X--Fe interaction. The substituted fission-product atom therefore does not replace the original Fe--Fe metallic bond with an equivalent cohesive bond. The TiN \(\Sigma 3\,[110](111)\) comparison in Fig.~\ref{fig:corresponding-lobster-cohp-coop}~(e) also shows some reduction in the direct X--N bonding metric relative to the clean Ti--N reference, shown by the black dashed curve. The penalty is smaller, \(-0.35~\mathrm{eV}\) per bond for Ce and \(-0.19~\mathrm{eV}\) per bond for Nd, and the occupied COOP response in Fig.~\ref{fig:corresponding-lobster-cohp-coop}~(f) remains largely positive. This agrees with the mechanical result: Nd and Ce weaken this unusually strong clean TiN boundary, but the decorated boundary still preserves a large absolute \(W_{\mathrm{sep}}\) and peak traction.

The \(\Sigma 11\,[110](332)\) comparison tests the same mechanism in a more complex boundary. In TiN, this boundary also has a much larger clean excess volume than \(\Sigma 3\,[110](111)\). In Fe, Fig.~\ref{fig:corresponding-lobster-cohp-coop}~(c) shows that the direct X--Fe bond again loses substantial occupied bonding intensity relative to the black dashed clean Fe--Fe reference, with \(\Delta\langle-\mathrm{ICOHP}\rangle_{\mathrm{occ}}=-0.87~\mathrm{eV}\) per bond for Ce and \(-0.80~\mathrm{eV}\) per bond for Nd. The corresponding Fe COOP curves in Fig.~\ref{fig:corresponding-lobster-cohp-coop}~(d) again indicate significant occupied antibonding character for the direct X--Fe bond. In TiN, Fig.~\ref{fig:corresponding-lobster-cohp-coop}~(g) shows that the X--N direct bond curve changes only weakly relative to the black dashed clean Ti--N reference, with \(\Delta\langle-\mathrm{ICOHP}\rangle_{\mathrm{occ}}=-0.07~\mathrm{eV}\) per bond for Ce and \(+0.09~\mathrm{eV}\) per bond for Nd, even though segregation to this TiN boundary is strongly favorable. The TiN COOP curves in Fig.~\ref{fig:corresponding-lobster-cohp-coop}~(h) also remain closer to the clean Ti--N response than the Fe curves in Fig.~\ref{fig:corresponding-lobster-cohp-coop}~(d). The \(\Sigma 11\) boundary gives a similar contrast. TiN can provide a strong segregation site for fission products while preserving the local bonding environment that carries GB strength, whereas Fe develops weak X--Fe bonding and reduced GB strength after segregation. Prior first-principles work on fission-product segregation in SiC showed that ceramic GBs can still be susceptible to fission-product-induced decohesion when segregation strongly disrupts the local bonding network \cite{enrique2014solute}. The TiN response observed here is therefore not a generic consequence of ceramic bonding, but reflects the ability of the surrounding Ti--N network to remain strong after Nd/Ce incorporation.

The bonding summary in Fig.~\ref{fig:lobster-bonding-summary}~(a) shows that this direct-bond pattern holds across the full set. Across all six sampled GBs, replacing an Fe atom with Nd or Ce produces consistently weaker X--Fe bonding than the clean Fe--Fe bonds at the substituted site, with a median \(\Delta\langle-\mathrm{ICOHP}\rangle_{\mathrm{occ}}\) of approximately \(-0.81~\mathrm{eV}\) per bond. The COOP sign check gives a consistent result. For every analyzed bcc Fe direct replacement case, including both Ce and Nd, the cumulative X--Fe integrated COOP (ICOOP) averaged over bonds at \(E_F\) is negative (Supplementary Table S3), indicating a net negative overlap population in the occupied states of the direct X--Fe bond.

TiN shows a different bonding response in Fig.~\ref{fig:lobster-bonding-summary}~(a). The X--N replacement bonds are much less degraded, with a median change of approximately \(-0.09~\mathrm{eV}\) per bond and several cases showing little change or slight strengthening. The cumulative X--N ICOOP averaged over bonds at \(E_F\) remains positive for all analyzed TiN direct replacement cases (Supplementary Table S3). The preserved local network in Fig.~\ref{fig:lobster-bonding-summary}~(b) shows the complementary trend, with Ti--N bonds near the segregant remaining nearly unchanged and a median \(\Delta\langle-\mathrm{ICOHP}\rangle_{\mathrm{occ}}\) close to zero. The supplementary ICOBI analysis provides a check on bond order. The ICOBI changes are much smaller than the COHP changes, indicating that the segregation-induced strength loss is not simply a loss of shared-electron population or bond order (Supplementary Table S2). The main contrast lies in the surrounding network. In bcc Fe, Nd/Ce substitution interrupts the local Fe--Fe bonding path at the GB and replaces it with weak X--Fe interactions that carry significant occupied antibonding character. In TiN, the fission-product atom changes the substituted site, but the surrounding Ti--N framework remains largely intact. The mixed ionic--covalent Ti--N network can therefore continue to carry much of the GB strength, making segregation less effective at producing decohesion.
\FloatBarrier
\section{Discussion}
The Fe/TiN contrast suggests that TiN coatings may contribute to FCCI mitigation in two ways. Much of the current strategy focuses on slowing fission-product transport before lanthanides reach the cladding interior. Refractory alloys have long been considered for advanced nuclear applications \cite{el2005review,murty2008structural,moschetti2022design}, and first-principles diffusion studies in bcc Fe and refractory bcc Mo show that lanthanide mobility can be lower in refractory metals \cite{ke2025diffusion,ke2026atomistic}. Recent TiN-coating studies also show that TiN layers can suppress Ce--Fe interaction in diffusion couples \cite{gan2024summary,gan2024effect}. The present calculations address another requirement for coating durability: if fission products reach an internal interface, the coating GB should avoid a large loss of strength.

Avoiding such strength loss is directly related to fission-product-induced intergranular weakening \cite{wang2022transmission,matthews2017fuel,keiser2019fuel}, which may contribute to environmentally assisted fracture and corrosion-cracking susceptibility in coated cladding systems. For coated cladding exposed to fission products, GBs should therefore be considered as diffusion pathways and as possible weak interfaces after fission-product segregation. The present calculations separate these two parts of the problem. Segregation energies describe where fission products prefer to reside, decohesion metrics describe the mechanical penalty of that residence, and bonding analysis identifies whether the local network that carries interfacial strength is disrupted or preserved. Although no high-throughput screening is performed here, this combination of segregation, decohesion, and COHP/COOP descriptors can be used as a future screening strategy for candidate nitride coatings when paired with transport, corrosion thermodynamics, GB character distributions, or continuum fracture models.

The present calculations focus on the intrinsic strength response of idealized CSL GBs under dilute, single-fission-product decoration. The GBs considered here are atomically well-defined and represent a limited subset of boundary structures. The purpose is to compare trends across Fe and TiN under identical cell construction and decohesion procedures, rather than to provide fully converged fracture energies for arbitrary GB structures. Each decorated supercell also contains only one Nd or Ce atom at its preferred segregation site. Effects associated with finite fission-product coverage, fission-product interactions, and co-segregation between fission products and pre-existing solute or impurity species in the cladding or coating are not included. At higher coverage, site competition and fission-product interactions could change the preferred segregation sites and the strength response.

The response of a real microstructure may also depend on the GB character distribution in each material. Ferritic and ferritic--martensitic steels can contain a mixture of low-\(\Sigma\) CSL and random high-angle boundaries, with distinct segregation, precipitation, and diffusion behavior during aging or under extreme conditions \cite{beladi2013relative,gupta2007role, wang2022transmission}. TiN coatings are often nanocrystalline and columnar, with GB networks that include both high-angle boundaries and boundaries at the atomic scale whose structures depend strongly on growth conditions \cite{sundgren1985structure,li2006grain}. These boundaries may range from open or voided interfaces to dense and coherent boundaries at the atomic scale that either facilitate or impede impurity diffusion. The present CSL-based comparison gives an atomistic baseline for selected boundary types, but does not weight the GB characters that appear in real coatings.

The calculations are also performed within a static, zero temperature DFT description. Diffusion kinetics, segregation dynamics under flux, irradiation-enhanced transport, corrosion reactions, and coupled chemo-mechanical processes are not modeled. Likewise, the decohesion analysis is restricted to idealized normal separation without crack-tip plasticity, dislocation emission, alternative fracture paths, or mixed mode loading. These effects may change absolute strengths, although the Fe/TiN comparison here uses the same loading and relaxation conditions for all cases. At finite temperature, multiple competing segregation configurations, site occupancies, and entropic contributions may also affect the thermodynamic response, particularly at higher fission-product concentrations \cite{spitaler2025ab,hu2024computational,lejcek2021entropy}. These effects could be treated with statistical-mechanical approaches, such as cluster expansion combined with Monte Carlo or mean-field methods \cite{ke2025role,van2004thermodynamics}.

Within these limits, the overall result shows that fission-product segregation and GB strength loss are not equivalent in TiN. TiN does not simply exclude Nd and Ce from internal interfaces, but segregated Nd and Ce cause a smaller strength penalty than in bcc Fe because the surrounding Ti--N network is largely preserved. These results support evaluating TiN and other nitride coatings by both fission-product transport and interfacial strength after segregation. The fitted traction--separation relations can also provide input for future cohesive-zone fracture models that extend the DFT decohesion response to larger length scales.

\section{Methods}
\label{sec:methods}
\subsection{Density functional theory calculations}
All first-principles calculations were performed within DFT using the Vienna Ab initio Simulation Package (VASP) \cite{PhysRevB.47.558,PhysRevB.54.11169}. The calculation settings followed related alloy studies \cite{KE201930,ke2022first,ke2025role}. Projector augmented-wave (PAW) potentials described ion-core and valence-electron interactions \cite{PhysRevB.50.17953,PhysRevB.59.1758}. Exchange--correlation effects were treated with the Perdew--Burke--Ernzerhof (PBE) generalized gradient approximation \cite{PhysRevLett.77.3865}. Spin polarization was included in all calculations, with ferromagnetic ordering initialized for bcc Fe.

A plane-wave cutoff energy of \(400~\mathrm{eV}\) was used for Fe-based systems, and \(550~\mathrm{eV}\) was used for TiN. Brillouin-zone integrations were performed using Monkhorst--Pack \(k\)-point meshes with target \(k\)-points-per-reciprocal-atom (KPPRA) values between 3000 and 4000. The electronic self-consistency tolerance was \(10^{-6}~\mathrm{eV}\), and atomic positions were relaxed until residual forces were below \(0.01~\mathrm{eV}\,\mathring{\mathrm{A}}^{-1}\). All structures were treated under periodic boundary conditions.

\subsection{Grain-boundary models}
\label{sec:gb-models}

Periodic GB supercells were constructed for both bcc Fe and TiN, with the GB plane normal oriented along the cell \(z\)-direction. Six CSL boundaries were selected to sample a range of interface structures: \(\Sigma 3\,[110](111)\), \(\Sigma 3\,[110](112)\), \(\Sigma 5\,[001](210)\), \(\Sigma 5\,[001](310)\), \(\Sigma 9\,[110](221)\), and \(\Sigma 11\,[110](332)\). The same nominal CSL set was used for Fe and TiN to make the material comparison direct.

The DFT GB configuration is illustrated in Fig.~\ref{fig:dft-gb-configuration}. The primitive CSL GB supercells were controlled to contain fewer than 100 atoms and were treated with periodic boundary conditions in all directions. Along the GB-normal direction, the supercell dimension was selected so that the separation between neighboring periodic GB images was greater than approximately \(10~\mathring{\mathrm{A}}\), reducing artificial interactions between periodic GBs.

Clean GB structures were fully relaxed before segregation and decohesion calculations. Nd and Ce were introduced as substitutional fission-product atoms at candidate sites at the GB, and the lowest-energy segregated configuration for each material, GB, and fission product is referred to as the best-site configuration. In Fe, Nd and Ce substituted for Fe atoms at the GB. In TiN, Nd and Ce were considered as substitutional fission-product species on Ti-sublattice sites near the GB, consistent with their metallic character and size mismatch relative to N. The number of symmetry-distinct candidate substitutional sites sampled for each GB is listed in Table~\ref{tab:gb-substitution-sites}. The same set of candidate sites was used for Nd and Ce within a given material and GB. When multiple site variants were sampled, all variants were included in the traction-curve figures and underlying dataset, while summary tables for best sites report only the variant with the lowest \(E_{\mathrm{seg}}\).
For each listed candidate site, a separate decorated supercell was constructed by replacing one host atom with one Nd or Ce atom. Multiple listed sites therefore correspond to separate configurations with one fission product, not simultaneous decoration at multiple sites.

\begin{table}[t]
\centering
\caption{Number of symmetry-distinct candidate GB substitutional sites sampled for Nd and Ce segregation. TiN site counts refer to Ti-sublattice sites.}
\label{tab:gb-substitution-sites}
\setlength{\tabcolsep}{8pt}
\tablefontsize
\begin{tabular}{lcc}
\toprule
GB & Fe sites & TiN Ti-sublattice sites \\
\midrule
\(\Sigma 3\,[110](111)\) & 1 & 1 \\
\(\Sigma 3\,[110](112)\) & 1 & 1 \\
\(\Sigma 5\,[001](210)\) & 2 & 1 \\
\(\Sigma 5\,[001](310)\) & 1 & 2 \\
\(\Sigma 9\,[110](221)\) & 3 & 2 \\
\(\Sigma 11\,[110](332)\) & 2 & 2 \\
\bottomrule
\end{tabular}
\end{table}

For decohesion calculations, a series of normal openings was imposed from \(0\) to \(6~\mathring{\mathrm{A}}\) in increments of \(0.10~\mathring{\mathrm{A}}\). At each opening, the out-of-plane lattice parameter was increased while the in-plane lattice vectors were kept fixed. The paired decohering and homogeneously strained reference configurations followed the \textit{ab initio} tensile-test approach of Enrique and Van der Ven~\cite{van2004thermodynamics,enrique2014solute,enrique2017decohesion}. In the decohering configurations, the opening was applied across the largest interlayer gap within \(4~\mathring{\mathrm{A}}\) of the GB center, so that separation occurred between atomic layers rather than through an atomic layer. Atomic relaxation was restricted to a slab centered on the GB and containing the three atomic layers nearest the interface, and atoms outside this slab were fixed. The same relaxed atom set was used in the decohering and homogeneously strained configurations. In the homogeneously strained reference configurations, the imposed elongation was distributed over the local elastic region within \(4~\mathring{\mathrm{A}}\) of the GB center. These paired calculations provide the total energy and elastic reference data used in Section~\ref{sec:decohesion-analysis} to construct the excess energy and traction--separation relationship.

The GB excess energy was calculated as
\begin{equation}
\gamma_{\mathrm{GB}} =
\frac{E_{\mathrm{GB}} - E_{\mathrm{ref}}}{2A},
\end{equation}
where \(E_{\mathrm{GB}}\) is the total energy of the relaxed GB supercell, \(E_{\mathrm{ref}}\) is the corresponding bulk reference energy, and \(A\) is the GB area. The factor of two accounts for the two periodic GBs in the supercell.

\subsection{Segregation energy}
The segregation energy of a fission-product atom \(X\), where \(X\) is Nd or Ce, was calculated as
\begin{equation}
E_{\mathrm{seg}} =
E_{\mathrm{GB}+X} - E_{\mathrm{GB}}
- \left(E_{\mathrm{bulk}+X} - E_{\mathrm{bulk}}\right),
\end{equation}
where \(E_{\mathrm{GB}+X}\) and \(E_{\mathrm{GB}}\) are the total energies of the GB system with and without fission product \(X\), and \(E_{\mathrm{bulk}+X}\) and \(E_{\mathrm{bulk}}\) are the corresponding bulk reference energies. A negative \(E_{\mathrm{seg}}\) indicates favorable segregation to the GB.

For comparisons normalized by area, the local segregation contribution used in the segregation-tolerance maps was written as
\begin{equation}
\Delta\gamma_{\mathrm{seg}}^{\mathrm{local}} =
\frac{E_{\mathrm{seg}}}{A}.
\label{eq:delta-gamma-seg}
\end{equation}
Here \(A\) is the GB area. This local normalization reports the segregation-energy change associated with the decorated interface. It differs by a factor of two from an average over the two periodic GB interfaces in the supercell. Because favorable segregation gives negative \(E_{\mathrm{seg}}\) and negative \(\Delta\gamma_{\mathrm{seg}}^{\mathrm{local}}\) in Eq.~\eqref{eq:delta-gamma-seg}, the segregation-tolerance maps in the Results plot the positive quantity \(-\Delta\gamma_{\mathrm{seg}}^{\mathrm{local}}\), referred to as the local reduction in segregation energy.

\subsection{Decohesion analysis}
\label{sec:decohesion-analysis}
The direct work of separation was used to quantify the energetic resistance of a GB to ideal cleavage into separated fracture surfaces:
\begin{equation}
W_{\mathrm{sep}} =
\frac{E_{\mathrm{sep}} - E_{\mathrm{GB}}}{A}.
\label{eq:wsep}
\end{equation}
Here \(E_{\mathrm{sep}}\) is the energy after separation into two free surfaces, and \(E_{\mathrm{GB}}\) is the energy of the intact GB configuration. The \(W_{\mathrm{sep}}\) values reported in the Results are the direct DFT energy differences at the largest calculated separation and represent the energy scale for full GB cleavage.

The traction--separation relationship was constructed from the paired decohering and homogeneously strained reference configurations described above, following the \textit{ab initio} tensile-test framework of Enrique and Van der Ven \cite{van2004thermodynamics,enrique2017decohesion}. For TiN, the stoichiometry of the periodic supercell was kept fixed throughout the separation path. Possible alternative Ti-rich or N-rich fracture surfaces were not sampled. The homogeneous reference was used to subtract the elastic strain energy stored in the non-decohering atomic layers adjacent to the opening plane. Its small-opening elastic response was fit as~\cite{enrique2014solute}
\begin{equation}
u_{\ell}(\delta_{\ell}) =
a\delta_{\ell}^2+b\delta_{\ell}^3+c\delta_{\ell}^4.
\end{equation}
A cohesive-zone description was then formulated in terms of thermodynamic excess quantities associated with the decohering GB plane. Because atomic relaxation was allowed in a finite slab centered on the GB, the total supercell energy \(u_T\) and total opening \(\delta_T\) include both the decohesion contribution and the elastic deformation of adjacent non-decohering planes. Following Enrique and Van der Ven~\cite{enrique2014solute,enrique2017decohesion}, the elastic contribution was evaluated from the homogeneous reference at the same normal stress:
\begin{align}
u_{\mathrm{ex}} &=
u_T - u_{\ell}\!\left[\delta_{\ell}\!\left(\sigma_T(\delta_T)\right)\right], \\
\delta_{\mathrm{ex}} &=
\delta_T - \delta_{\ell}\!\left[\sigma_T(\delta_T)\right].
\end{align}
Here \(u_{\ell}\) and \(\delta_{\ell}\) are the energy and elongation of the adjacent elastic region when it is part of a homogeneous solid at the same normal stress \(\sigma_T\) as the decohering supercell. The equivalent elastic opening \(\delta_{\ell}\) was obtained by matching the normal force from the decohering calculation to \(\mathrm{d}u_{\ell}/\mathrm{d}\delta_{\ell}\). In the discrete-layer implementation, the homogeneous reference was evaluated for the same slab geometry centered on the GB and scaled according to the number of relaxable atomic layers so that finite size effects from the constrained relaxation region were treated consistently between decohering and homogeneous calculations.

The resulting excess energy was fit using the Rose universal binding-energy relationship (UBER) in the form used by Enrique and Van der Ven~\cite{rose1981universal,enrique2017decohesion}:
\begin{equation}
u_{\mathrm{ex}}(\delta_{\mathrm{ex}})=
u_0+2\gamma_{\mathrm{c}}\left[1-\left(1+\frac{\delta_{\mathrm{ex}}}{\lambda}\right)
\exp\left(-\frac{\delta_{\mathrm{ex}}}{\lambda}\right)\right],
\end{equation}
where \(u_0\) is the fitted zero-opening offset, \(2\gamma_{\mathrm{c}}\) is the fitted energy scale, and \(\lambda\) is a characteristic separation length. For an ideal excess-energy reference, \(u_0\) is expected to be close to zero. The traction was obtained from
\begin{equation}
T(\delta_{\mathrm{ex}})=
\frac{1}{A}\frac{\mathrm{d}u_{\mathrm{ex}}}{\mathrm{d}\delta_{\mathrm{ex}}},
\end{equation}
and the maximum of this curve defines the cohesive strength \(\sigma_{\max}\). In the present ideal tensile geometry, \(\sigma_{\max}\) is the maximum cohesive stress along the ideal opening path and provides an atomistic measure of resistance to the onset of local separation.

The UBER fits were used only to obtain smooth traction--separation curves and peak tractions. The reported \(W_{\mathrm{sep}}\) values, changes in \(W_{\mathrm{sep}}\), and retention values based on \(W_{\mathrm{sep}}\) were taken from direct DFT total-energy differences.

\subsection{Chemical-bonding analysis}
\label{sec:chemical-bonding-analysis}
Chemical bonding was analyzed using crystal-orbital Hamilton population (COHP)~\cite{dronskowski1993crystal} and crystal-orbital overlap population (COOP)~\cite{hughbanks1983chains} calculations, which were performed using LOBSTER~\cite{maintz2016lobster,nelson2020lobster}. These static post-processing calculations used relaxed clean and best-site decorated GB structures. The analysis was used only to interpret bonding trends and was not used to define \(W_{\mathrm{sep}}\), \(\sigma_{\max}\), or the retention metrics.

The LOBSTER calculations used the pbeVaspFit2015 basis set. The local basis functions were Fe \(3d\,4s\,4p\), Ti \(3s\,3p\,3d\,4s\), N \(2s\,2p\), and Ce/Nd \(5s\,5p\,5d\,6s\), consistent with the PAW datasets used for the static calculations. COHP and COOP curves were generated for atom pairs within \(3.4~\mathring{\mathrm{A}}\) in Fe and \(3.2~\mathring{\mathrm{A}}\) in TiN. Projection quality was monitored using the LOBSTER charge spilling and recovered electron count, and the same basis and pair-selection rules were used for clean and decorated structures within each GB. The bonding analysis was performed for all six sampled GBs using the relaxed clean structures and the best-site Nd- and Ce-decorated structures for Fe and TiN.

Two local bond groups were compared. The first tests whether the fission-product atom replaces the original matrix bond with an equally strong bond, while the second tests whether nearby Fe--Fe or Ti--N bonds survive around that atom. The direct replacement bond compares bonds between X and its neighboring atoms in the decorated GB with the corresponding clean bonds between the matrix atom and its neighboring atoms at the substituted site: X--Fe relative to Fe--Fe in bcc Fe and X--N relative to Ti--N in TiN, where X is Nd or Ce. The preserved local network compares nearby matrix-neighbor bonds that do not contain X: Fe--Fe bonds near X in Fe and Ti--N bonds near X in TiN. The direct bond group included X--Fe pairs within \(3.4~\mathring{\mathrm{A}}\) in Fe and X--N pairs within \(3.2~\mathring{\mathrm{A}}\) in TiN. The network group included Fe--Fe pairs within \(3.0~\mathring{\mathrm{A}}\) or Ti--N pairs within \(2.45~\mathring{\mathrm{A}}\) whose atoms or bond midpoint were within \(4.0~\mathring{\mathrm{A}}\) of the segregant. In both cases, the plotted curves and scalar metrics were averaged over all atom pairs satisfying the bond selection criteria based on distance. The same distance criteria were then applied to the corresponding clean structure for reference. For each bond group, the occupied \(-\mathrm{COHP}\) averaged over bonds and integrated over the energy window was defined as:
\begin{equation}
\langle-\mathrm{ICOHP}\rangle_{\mathrm{occ}} =
\frac{1}{N_b}\sum_{i=1}^{N_b}
\int_{-10~\mathrm{eV}}^{E_F}
\left[-\mathrm{COHP}_i(E)\right]\,\mathrm{d}E,
\label{eq:icohp-occ}
\end{equation}
where \(N_b\) is the number of atom pairs in the selected bond group. Larger positive \(\langle-\mathrm{ICOHP}\rangle_{\mathrm{occ}}\) values in Eq.~\eqref{eq:icohp-occ} indicate stronger average bonding over the occupied states included in this energy range. The subscript ``occ'' denotes integration over occupied states up to \(E_F\). The lower bound of \(-10~\mathrm{eV}\) was chosen to include the main occupied valence-bonding states while avoiding deeper semicore contributions. Extending the lower bound to \(-12~\mathrm{eV}\) gave essentially the same differential bonding trends, as shown in Supplementary Figure S1 and Supplementary Table S1, whereas ICOHP values over the full LOBSTER integration range include deeper states that can obscure the valence-bonding changes associated with segregation. In practice, this quantity was evaluated from the cumulative ICOHP curves written by LOBSTER by subtracting the value at \(-10~\mathrm{eV}\) from the value at \(E_F\), after averaging over the selected atom pairs. Changes caused by segregation were then calculated as:
\begin{equation}
\Delta\langle-\mathrm{ICOHP}\rangle_{\mathrm{occ}} =
\langle-\mathrm{ICOHP}\rangle_{\mathrm{occ}}^{\mathrm{decorated}} -
\langle-\mathrm{ICOHP}\rangle_{\mathrm{occ}}^{\mathrm{clean}}.
\label{eq:delta-icohp-occ}
\end{equation}
Positive values indicate increased bonding relative to the corresponding clean bond group, while negative values indicate reduced bonding.

Energy-resolved COHP and COOP curves were averaged over all atom pairs in the selected bond group. In the plotted convention, positive \(-\mathrm{COHP}\) corresponds to bonding contributions, whereas negative COOP indicates antibonding overlap population. The scalar COOP values in Supplementary Table S3 are cumulative LOBSTER ICOOP values at \(E_F\), averaged over the selected bond group, and were used only as a sign check rather than as a metric integrated from \(-10~\mathrm{eV}\) to \(E_F\). ICOBI values were also evaluated as a supplementary measure of bond order~\cite{muller2021cobi} for the same representative GBs, with the complete values listed in Supplementary Table S2. For the energy-resolved comparison, best-site Ce- and Nd-decorated \(\Sigma 3\,[110](111)\) and \(\Sigma 11\,[110](332)\) GBs were selected to compare an ordered low-\(\Sigma\) boundary with a higher-index boundary.

\backmatter

\section*{Data availability}
The data that support the findings of this study are available from the corresponding authors upon reasonable request.

\section*{Code availability}
No new custom software or algorithm was developed for this study. The analysis scripts used to process the first-principles results and generate the figures are available from the corresponding authors upon reasonable request.

\section*{Acknowledgements}
This research was supported by the Innovative Nuclear Materials (INM) Program under the U.S. Department of Energy, Office of Nuclear Energy. This manuscript has been authored by Battelle Energy Alliance, LLC under Contract No.~DE-AC07-05ID14517 with the U.S. Department of Energy. The United States Government retains, and the publisher, by accepting the article for publication, acknowledges that the United States Government retains a nonexclusive, paid-up, irrevocable, world-wide license to publish or reproduce the published form of this manuscript, or allow others to do so, for United States Government purposes. This research made use of the resources of the High Performance Computing Center at Idaho National Laboratory, which is supported by the Office of Nuclear Energy of the U.S. Department of Energy and the Nuclear Science User Facilities under Contract No. DE-AC07-05ID14517.

\section*{Author contributions}
J.-H.K.: conceptualization, formal analysis, investigation, methodology (lead), visualization, data curation, and writing--original draft. R.S.: conceptualization, investigation, methodology, funding acquisition, supervision, and writing--review and editing.

\section*{Competing interests}
The authors declare no competing interests.

\bibliography{sn-bibliography}

\clearpage
\setcounter{section}{0}
\renewcommand{\thesection}{S\arabic{section}}
\setcounter{figure}{0}
\renewcommand{\thefigure}{S\arabic{figure}}
\setcounter{table}{0}
\renewcommand{\thetable}{S\arabic{table}}
\setcounter{equation}{0}
\renewcommand{\theequation}{S\arabic{equation}}

\section*{Supplementary Information}
\addcontentsline{toc}{section}{Supplementary Information}

\section{Sensitivity of the window-integrated bonding metric}

The main text uses an occupied-state crystal-orbital Hamilton population (COHP) metric from the Local Orbital Basis Suite Toward Electronic-Structure Reconstruction (LOBSTER)~\cite{dronskowski1993crystal,maintz2016lobster,nelson2020lobster}. The integrated value is denoted ICOHP.
For a selected bond group, the quantity integrated over the energy window is
\begin{equation}
\langle-\mathrm{ICOHP}\rangle_{\mathrm{occ}} =
\frac{1}{N_b}\sum_{i=1}^{N_b}
\int_{-10~\mathrm{eV}}^{E_F}
\left[-\mathrm{COHP}_i(E)\right]\,\mathrm{d}E.
\end{equation}
This quantity was used to compare local bonding changes caused by Nd and Ce
segregation. The lower bound of
\(-10~\mathrm{eV}\) was chosen to include the main occupied valence-bonding states
while excluding deeper states that do not control the bonding changes discussed in
the manuscript.
Changes in this quantity are reported as decorated minus corresponding clean
bond groups. Negative \(\Delta\langle-\mathrm{ICOHP}\rangle_{\mathrm{occ}}\)
therefore indicates a loss of occupied bonding stabilization, whereas positive
values indicate stronger occupied bonding over the selected energy window.

\begin{figure}[h]
\centering
\includegraphics[width=0.82\textwidth]{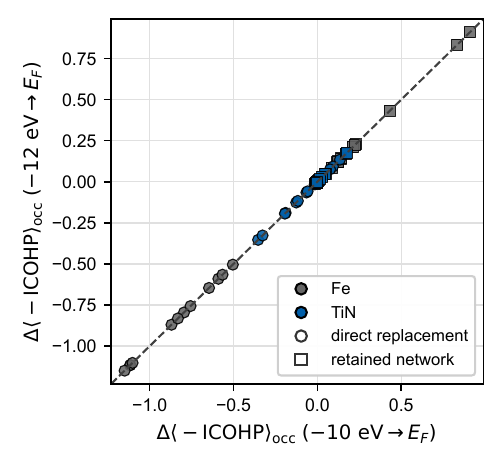}
\caption{Sensitivity of \(\Delta\langle-\mathrm{ICOHP}\rangle_{\mathrm{occ}}\)
to the lower integration bound. The horizontal axis uses the
\(-10~\mathrm{eV}\rightarrow E_F\) window used in the main text, and the vertical
axis uses \(-12~\mathrm{eV}\rightarrow E_F\). Each point is one material, solute,
GB, and bond group. The dashed line has unit slope.}
\label{fig:si-window-sensitivity}
\end{figure}

To check the sensitivity of this choice, the same differential metric was recomputed
using lower bounds of \(-8~\mathrm{eV}\) and \(-12~\mathrm{eV}\). Figure~\ref{fig:si-window-sensitivity}
compares the \(-10~\mathrm{eV}\rightarrow E_F\) and
\(-12~\mathrm{eV}\rightarrow E_F\) values for all analyzed bond groups, including
all six grain boundaries (GBs), both solutes, both materials, and the direct replacement and preserved
network bond groups. The data lie very close to the line of unit slope, showing that
the reported trends do not depend on the exact lower bound within this range.

The values in Table~\ref{tab:si-window-sensitivity} give the same result for
each GB and solute. The \(-10~\mathrm{eV}\) and \(-12~\mathrm{eV}\) values are
nearly identical, and comparison with the \(-8~\mathrm{eV}\) values gives the
same direct bond and preserved network trends. For this reason, the manuscript
uses integration from \(-10~\mathrm{eV}\) to \(E_F\) to focus on
the valence-bonding changes associated with segregation. Overall, the table
shows the same trend emphasized in the main text: in Fe, replacing Fe with Nd or
Ce gives negative \(\Delta\langle-\mathrm{ICOHP}\rangle_{\mathrm{occ}}\) values
for the direct \(X\)--Fe bonds. These negative values show that the replacement
bonds lose substantial occupied bonding and are consistently weaker than the
corresponding clean Fe--Fe bonds at the substituted site. In TiN, the \(X\)--N
replacement bonds are less degraded, and the preserved Ti--N network remains
close to the clean reference.

\begin{center}
\footnotesize
\setlength{\tabcolsep}{3pt}
\begin{longtable}{lllrrrr}
\caption{\(\Delta\langle-\mathrm{ICOHP}\rangle_{\mathrm{occ}}\) values for
different lower integration bounds. Values are in eV per bond. Direct denotes
the direct replacement bond group, and the Retained row label denotes the
preserved local network bond group.}
\label{tab:si-window-sensitivity}\\
\toprule
Material & GB & Solute & Group &
\(-8~\mathrm{eV}\) & \(-10~\mathrm{eV}\) & \(-12~\mathrm{eV}\) \\
\midrule
\endfirsthead
\toprule
Material & GB & Solute & Group &
\(-8~\mathrm{eV}\) & \(-10~\mathrm{eV}\) & \(-12~\mathrm{eV}\) \\
\midrule
\endhead
Fe & \(\Sigma 3\,[110](111)\) & Ce & Direct & -1.116 & -1.116 & -1.116 \\
Fe & \(\Sigma 3\,[110](111)\) & Ce & Retained & +0.014 & +0.011 & +0.011 \\
Fe & \(\Sigma 3\,[110](111)\) & Nd & Direct & -1.116 & -1.116 & -1.116 \\
Fe & \(\Sigma 3\,[110](111)\) & Nd & Retained & -0.013 & -0.016 & -0.016 \\
Fe & \(\Sigma 3\,[110](112)\) & Ce & Direct & -1.152 & -1.149 & -1.149 \\
Fe & \(\Sigma 3\,[110](112)\) & Ce & Retained & +0.121 & +0.123 & +0.123 \\
Fe & \(\Sigma 3\,[110](112)\) & Nd & Direct & -1.105 & -1.102 & -1.102 \\
Fe & \(\Sigma 3\,[110](112)\) & Nd & Retained & +0.140 & +0.142 & +0.142 \\
Fe & \(\Sigma 5\,[001](210)\) & Ce & Direct & -0.833 & -0.832 & -0.832 \\
Fe & \(\Sigma 5\,[001](210)\) & Ce & Retained & +0.228 & +0.228 & +0.228 \\
Fe & \(\Sigma 5\,[001](210)\) & Nd & Direct & -0.758 & -0.756 & -0.756 \\
Fe & \(\Sigma 5\,[001](210)\) & Nd & Retained & +0.211 & +0.211 & +0.211 \\
Fe & \(\Sigma 5\,[001](310)\) & Ce & Direct & -0.599 & -0.590 & -0.590 \\
Fe & \(\Sigma 5\,[001](310)\) & Ce & Retained & +0.904 & +0.909 & +0.909 \\
Fe & \(\Sigma 5\,[001](310)\) & Nd & Direct & -0.513 & -0.504 & -0.504 \\
Fe & \(\Sigma 5\,[001](310)\) & Nd & Retained & +0.828 & +0.834 & +0.834 \\
Fe & \(\Sigma 9\,[110](221)\) & Ce & Direct & -0.658 & -0.646 & -0.646 \\
Fe & \(\Sigma 9\,[110](221)\) & Ce & Retained & +0.402 & +0.432 & +0.432 \\
Fe & \(\Sigma 9\,[110](221)\) & Nd & Direct & -0.576 & -0.566 & -0.566 \\
Fe & \(\Sigma 9\,[110](221)\) & Nd & Retained & +0.200 & +0.222 & +0.222 \\
Fe & \(\Sigma 11\,[110](332)\) & Ce & Direct & -0.870 & -0.870 & -0.870 \\
Fe & \(\Sigma 11\,[110](332)\) & Ce & Retained & +0.121 & +0.120 & +0.120 \\
Fe & \(\Sigma 11\,[110](332)\) & Nd & Direct & -0.795 & -0.795 & -0.795 \\
Fe & \(\Sigma 11\,[110](332)\) & Nd & Retained & +0.087 & +0.085 & +0.085 \\
TiN & \(\Sigma 3\,[110](111)\) & Ce & Direct & -0.352 & -0.353 & -0.353 \\
TiN & \(\Sigma 3\,[110](111)\) & Ce & Retained & +0.052 & +0.049 & +0.049 \\
TiN & \(\Sigma 3\,[110](111)\) & Nd & Direct & -0.188 & -0.190 & -0.190 \\
TiN & \(\Sigma 3\,[110](111)\) & Nd & Retained & +0.049 & +0.046 & +0.046 \\
TiN & \(\Sigma 3\,[110](112)\) & Ce & Direct & -0.194 & -0.194 & -0.194 \\
TiN & \(\Sigma 3\,[110](112)\) & Ce & Retained & +0.170 & +0.169 & +0.169 \\
TiN & \(\Sigma 3\,[110](112)\) & Nd & Direct & -0.058 & -0.058 & -0.058 \\
TiN & \(\Sigma 3\,[110](112)\) & Nd & Retained & +0.176 & +0.175 & +0.175 \\
TiN & \(\Sigma 5\,[001](210)\) & Ce & Direct & -0.327 & -0.327 & -0.327 \\
TiN & \(\Sigma 5\,[001](210)\) & Ce & Retained & +0.004 & +0.004 & +0.004 \\
TiN & \(\Sigma 5\,[001](210)\) & Nd & Direct & -0.126 & -0.126 & -0.126 \\
TiN & \(\Sigma 5\,[001](210)\) & Nd & Retained & -0.001 & -0.001 & -0.001 \\
TiN & \(\Sigma 5\,[001](310)\) & Ce & Direct & -0.117 & -0.117 & -0.117 \\
TiN & \(\Sigma 5\,[001](310)\) & Ce & Retained & +0.030 & +0.030 & +0.030 \\
TiN & \(\Sigma 5\,[001](310)\) & Nd & Direct & +0.075 & +0.075 & +0.075 \\
TiN & \(\Sigma 5\,[001](310)\) & Nd & Retained & +0.015 & +0.014 & +0.014 \\
TiN & \(\Sigma 9\,[110](221)\) & Ce & Direct & -0.016 & -0.016 & -0.016 \\
TiN & \(\Sigma 9\,[110](221)\) & Ce & Retained & -0.006 & -0.007 & -0.007 \\
TiN & \(\Sigma 9\,[110](221)\) & Nd & Direct & +0.134 & +0.134 & +0.134 \\
TiN & \(\Sigma 9\,[110](221)\) & Nd & Retained & -0.004 & -0.005 & -0.005 \\
TiN & \(\Sigma 11\,[110](332)\) & Ce & Direct & -0.065 & -0.065 & -0.065 \\
TiN & \(\Sigma 11\,[110](332)\) & Ce & Retained & -0.004 & -0.004 & -0.004 \\
TiN & \(\Sigma 11\,[110](332)\) & Nd & Direct & +0.093 & +0.093 & +0.093 \\
TiN & \(\Sigma 11\,[110](332)\) & Nd & Retained & -0.005 & -0.005 & -0.005 \\
\bottomrule
\end{longtable}
\end{center}

\section{Supplementary bond-index analysis}

Crystal-orbital bond index (COBI) analysis~\cite{muller2021cobi} was used as a secondary check on the
bonding interpretation. In LOBSTER, the integrated COBI (ICOBI) is a
dimensionless measure related to bond order that reflects the amount of shared
electron population associated with a selected atom pair. Larger positive ICOBI
values indicate stronger bond order, but ICOBI is not an energetic
measure of bond strength. It is therefore complementary to the COHP and crystal-orbital overlap population (COOP) analysis
used in the main text.

\begin{figure}[h]
\centering
\includegraphics[width=0.95\textwidth]{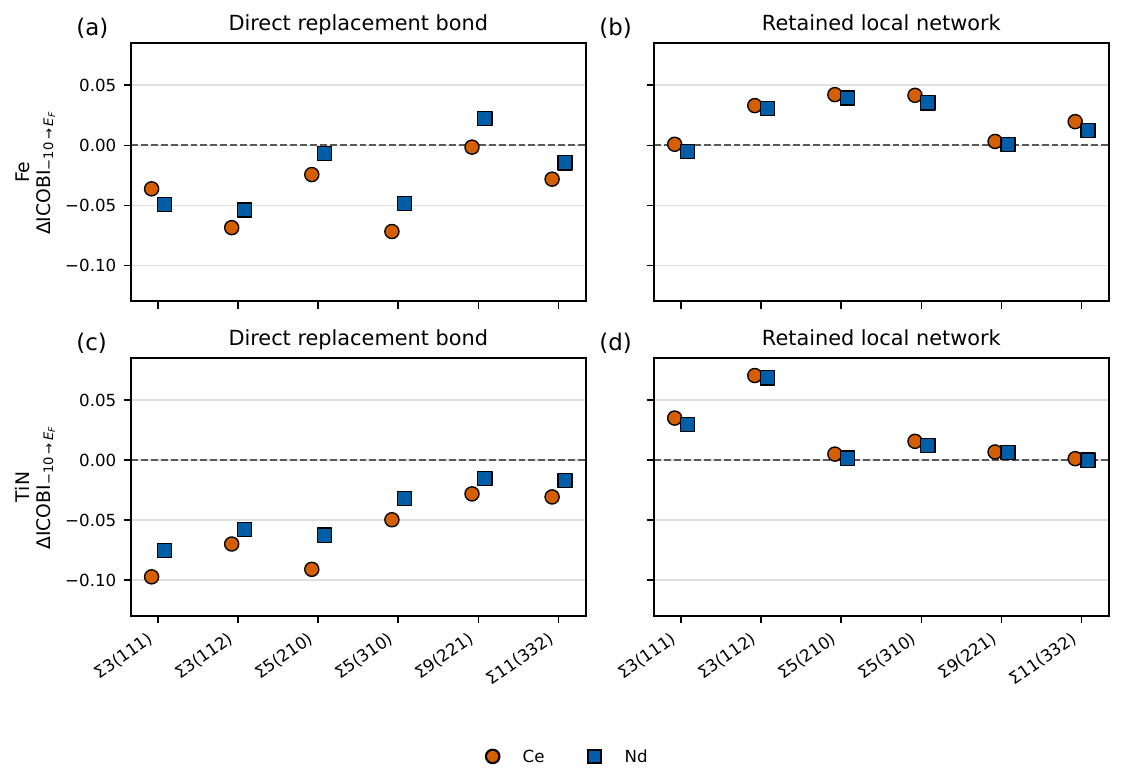}
\caption{ICOBI changes integrated over the energy window for all six sampled GBs using the
\(-10~\mathrm{eV}\rightarrow E_F\) energy range. (a) and (b) show direct
replacement bonds and preserved local network bonds in Fe, respectively. (c) and
(d) show the corresponding quantities in TiN. Values are differences between
decorated and corresponding clean bond groups. Positive values indicate larger
ICOBI after segregation.}
\label{fig:si-icobi-summary}
\end{figure}

For consistency with the COHP analysis, Figure~\ref{fig:si-icobi-summary} compares
ICOBI changes integrated from \(-10~\mathrm{eV}\) to \(E_F\) for all six
GBs in body-centered cubic (bcc) Fe and TiN. The same direct replacement and preserved local network
bond groups were used. Table~\ref{tab:si-icobi-summary} gives the corresponding
values for lower bounds of \(-8\), \(-10\), and \(-12~\mathrm{eV}\). The ICOBI
changes are much smaller than the COHP changes in the main text, which indicates
that segregation-induced cohesion loss is not simply a collapse of bond order
character. As shown by the ICOHP analysis, the more important distinction is the
energetic quality of the replacement bond and the preservation of the
surrounding Ti--N network.

\begin{center}
\footnotesize
\setlength{\tabcolsep}{3pt}
\begin{longtable}{lllrrrr}
\caption{Window integrated \(\Delta\mathrm{ICOBI}\) values for different lower
integration bounds. ICOBI is dimensionless. Direct denotes the direct
replacement bond group, and the Retained row label denotes the preserved local
network bond group.}
\label{tab:si-icobi-summary}\\
\toprule
Material & GB & Solute & Group &
\(-8~\mathrm{eV}\) & \(-10~\mathrm{eV}\) & \(-12~\mathrm{eV}\) \\
\midrule
\endfirsthead
\toprule
Material & GB & Solute & Group &
\(-8~\mathrm{eV}\) & \(-10~\mathrm{eV}\) & \(-12~\mathrm{eV}\) \\
\midrule
\endhead
Fe & \(\Sigma 3\,[110](111)\) & Ce & Direct & -0.036 & -0.036 & -0.036 \\
Fe & \(\Sigma 3\,[110](111)\) & Ce & Retained & +0.001 & +0.001 & +0.001 \\
Fe & \(\Sigma 3\,[110](111)\) & Nd & Direct & -0.049 & -0.049 & -0.049 \\
Fe & \(\Sigma 3\,[110](111)\) & Nd & Retained & -0.005 & -0.005 & -0.005 \\
Fe & \(\Sigma 3\,[110](112)\) & Ce & Direct & -0.068 & -0.069 & -0.069 \\
Fe & \(\Sigma 3\,[110](112)\) & Ce & Retained & +0.034 & +0.033 & +0.033 \\
Fe & \(\Sigma 3\,[110](112)\) & Nd & Direct & -0.053 & -0.054 & -0.054 \\
Fe & \(\Sigma 3\,[110](112)\) & Nd & Retained & +0.031 & +0.031 & +0.031 \\
Fe & \(\Sigma 5\,[001](210)\) & Ce & Direct & -0.024 & -0.024 & -0.024 \\
Fe & \(\Sigma 5\,[001](210)\) & Ce & Retained & +0.042 & +0.042 & +0.042 \\
Fe & \(\Sigma 5\,[001](210)\) & Nd & Direct & -0.007 & -0.007 & -0.007 \\
Fe & \(\Sigma 5\,[001](210)\) & Nd & Retained & +0.040 & +0.040 & +0.040 \\
Fe & \(\Sigma 5\,[001](310)\) & Ce & Direct & -0.071 & -0.072 & -0.072 \\
Fe & \(\Sigma 5\,[001](310)\) & Ce & Retained & +0.042 & +0.042 & +0.042 \\
Fe & \(\Sigma 5\,[001](310)\) & Nd & Direct & -0.048 & -0.048 & -0.048 \\
Fe & \(\Sigma 5\,[001](310)\) & Nd & Retained & +0.036 & +0.035 & +0.035 \\
Fe & \(\Sigma 9\,[110](221)\) & Ce & Direct & -0.001 & -0.001 & -0.001 \\
Fe & \(\Sigma 9\,[110](221)\) & Ce & Retained & +0.004 & +0.003 & +0.003 \\
Fe & \(\Sigma 9\,[110](221)\) & Nd & Direct & +0.023 & +0.022 & +0.022 \\
Fe & \(\Sigma 9\,[110](221)\) & Nd & Retained & +0.001 & +0.001 & +0.001 \\
Fe & \(\Sigma 11\,[110](332)\) & Ce & Direct & -0.028 & -0.028 & -0.028 \\
Fe & \(\Sigma 11\,[110](332)\) & Ce & Retained & +0.020 & +0.020 & +0.020 \\
Fe & \(\Sigma 11\,[110](332)\) & Nd & Direct & -0.014 & -0.015 & -0.015 \\
Fe & \(\Sigma 11\,[110](332)\) & Nd & Retained & +0.013 & +0.013 & +0.013 \\
TiN & \(\Sigma 3\,[110](111)\) & Ce & Direct & -0.097 & -0.097 & -0.097 \\
TiN & \(\Sigma 3\,[110](111)\) & Ce & Retained & +0.035 & +0.035 & +0.035 \\
TiN & \(\Sigma 3\,[110](111)\) & Nd & Direct & -0.075 & -0.075 & -0.075 \\
TiN & \(\Sigma 3\,[110](111)\) & Nd & Retained & +0.030 & +0.029 & +0.029 \\
TiN & \(\Sigma 3\,[110](112)\) & Ce & Direct & -0.070 & -0.070 & -0.070 \\
TiN & \(\Sigma 3\,[110](112)\) & Ce & Retained & +0.070 & +0.070 & +0.070 \\
TiN & \(\Sigma 3\,[110](112)\) & Nd & Direct & -0.058 & -0.058 & -0.058 \\
TiN & \(\Sigma 3\,[110](112)\) & Nd & Retained & +0.069 & +0.068 & +0.068 \\
TiN & \(\Sigma 5\,[001](210)\) & Ce & Direct & -0.091 & -0.091 & -0.091 \\
TiN & \(\Sigma 5\,[001](210)\) & Ce & Retained & +0.005 & +0.005 & +0.005 \\
TiN & \(\Sigma 5\,[001](210)\) & Nd & Direct & -0.063 & -0.063 & -0.063 \\
TiN & \(\Sigma 5\,[001](210)\) & Nd & Retained & +0.002 & +0.002 & +0.002 \\
TiN & \(\Sigma 5\,[001](310)\) & Ce & Direct & -0.050 & -0.050 & -0.050 \\
TiN & \(\Sigma 5\,[001](310)\) & Ce & Retained & +0.016 & +0.016 & +0.016 \\
TiN & \(\Sigma 5\,[001](310)\) & Nd & Direct & -0.032 & -0.032 & -0.032 \\
TiN & \(\Sigma 5\,[001](310)\) & Nd & Retained & +0.012 & +0.012 & +0.012 \\
TiN & \(\Sigma 9\,[110](221)\) & Ce & Direct & -0.028 & -0.028 & -0.028 \\
TiN & \(\Sigma 9\,[110](221)\) & Ce & Retained & +0.007 & +0.007 & +0.007 \\
TiN & \(\Sigma 9\,[110](221)\) & Nd & Direct & -0.015 & -0.015 & -0.015 \\
TiN & \(\Sigma 9\,[110](221)\) & Nd & Retained & +0.006 & +0.006 & +0.006 \\
TiN & \(\Sigma 11\,[110](332)\) & Ce & Direct & -0.031 & -0.031 & -0.031 \\
TiN & \(\Sigma 11\,[110](332)\) & Ce & Retained & +0.001 & +0.001 & +0.001 \\
TiN & \(\Sigma 11\,[110](332)\) & Nd & Direct & -0.017 & -0.017 & -0.017 \\
TiN & \(\Sigma 11\,[110](332)\) & Nd & Retained & -0.000 & -0.000 & -0.000 \\
\bottomrule
\end{longtable}
\end{center}

\section{Supplementary overlap-population values for direct replacement bonds}

The main text uses COHP as the primary energetic bonding descriptor. Crystal-orbital overlap population (COOP)~\cite{hughbanks1983chains} was
used as a sign check for occupied bonding or antibonding overlap population in
the direct replacement bonds. Table~\ref{tab:si-direct-icoop} lists the
integrated COOP (ICOOP) values at the Fermi level averaged over bonds from the LOBSTER
output. These values are not restricted to the \(-10~\mathrm{eV}\) window used
for the COHP metric in the main text. The decorated X--Fe values are negative for all
analyzed bcc Fe cases, indicating net negative occupied overlap population for
the direct fission-product--Fe interaction. In contrast, the decorated X--N
values remain positive for all analyzed TiN cases.

\begin{center}
\footnotesize
\setlength{\tabcolsep}{3pt}
\begin{longtable}{llllrrr}
\caption{Cumulative ICOOP values at the Fermi level averaged over bonds for the direct
replacement bond groups. Values are dimensionless and are not integrated over the energy window
from \(-10~\mathrm{eV}\). X denotes Ce or Nd.}
\label{tab:si-direct-icoop}\\
\toprule
Material & GB & Solute & Bond &
\(\mathrm{ICOOP}_{E_F}^{\mathrm{clean}}\) &
\(\mathrm{ICOOP}_{E_F}^{\mathrm{decorated}}\) &
\(\Delta\mathrm{ICOOP}_{E_F}\) \\
\midrule
\endfirsthead
\toprule
Material & GB & Solute & Bond &
\(\mathrm{ICOOP}_{E_F}^{\mathrm{clean}}\) &
\(\mathrm{ICOOP}_{E_F}^{\mathrm{decorated}}\) &
\(\Delta\mathrm{ICOOP}_{E_F}\) \\
\midrule
\endhead
Fe & \(\Sigma 3\,[110](111)\) & Ce & X--Fe & +0.377 & -0.155 & -0.532 \\
Fe & \(\Sigma 3\,[110](111)\) & Nd & X--Fe & +0.377 & -0.228 & -0.605 \\
Fe & \(\Sigma 3\,[110](112)\) & Ce & X--Fe & +0.118 & -0.342 & -0.460 \\
Fe & \(\Sigma 3\,[110](112)\) & Nd & X--Fe & +0.118 & -0.460 & -0.578 \\
Fe & \(\Sigma 5\,[001](210)\) & Ce & X--Fe & +0.045 & -0.376 & -0.421 \\
Fe & \(\Sigma 5\,[001](210)\) & Nd & X--Fe & +0.045 & -0.971 & -1.016 \\
Fe & \(\Sigma 5\,[001](310)\) & Ce & X--Fe & -0.708 & -0.686 & +0.022 \\
Fe & \(\Sigma 5\,[001](310)\) & Nd & X--Fe & -0.708 & -0.627 & +0.081 \\
Fe & \(\Sigma 9\,[110](221)\) & Ce & X--Fe & -0.350 & -0.315 & +0.036 \\
Fe & \(\Sigma 9\,[110](221)\) & Nd & X--Fe & -0.350 & -0.277 & +0.074 \\
Fe & \(\Sigma 11\,[110](332)\) & Ce & X--Fe & +0.209 & -0.401 & -0.610 \\
Fe & \(\Sigma 11\,[110](332)\) & Nd & X--Fe & +0.209 & -0.593 & -0.802 \\
TiN & \(\Sigma 3\,[110](111)\) & Ce & X--N & +0.109 & +0.092 & -0.017 \\
TiN & \(\Sigma 3\,[110](111)\) & Nd & X--N & +0.109 & +0.118 & +0.009 \\
TiN & \(\Sigma 3\,[110](112)\) & Ce & X--N & +0.096 & +0.081 & -0.016 \\
TiN & \(\Sigma 3\,[110](112)\) & Nd & X--N & +0.096 & +0.095 & -0.001 \\
TiN & \(\Sigma 5\,[001](210)\) & Ce & X--N & +0.114 & +0.076 & -0.038 \\
TiN & \(\Sigma 5\,[001](210)\) & Nd & X--N & +0.114 & +0.091 & -0.023 \\
TiN & \(\Sigma 5\,[001](310)\) & Ce & X--N & +0.105 & +0.084 & -0.021 \\
TiN & \(\Sigma 5\,[001](310)\) & Nd & X--N & +0.105 & +0.096 & -0.009 \\
TiN & \(\Sigma 9\,[110](221)\) & Ce & X--N & +0.081 & +0.084 & +0.003 \\
TiN & \(\Sigma 9\,[110](221)\) & Nd & X--N & +0.081 & +0.099 & +0.019 \\
TiN & \(\Sigma 11\,[110](332)\) & Ce & X--N & +0.077 & +0.071 & -0.006 \\
TiN & \(\Sigma 11\,[110](332)\) & Nd & X--N & +0.077 & +0.085 & +0.008 \\
\bottomrule
\end{longtable}
\end{center}

\end{document}